\documentclass[11pt]{article}
\usepackage{verbatim,amsmath,amssymb}
\usepackage{epsfig,float,color}
\usepackage{epstopdf}
\usepackage{geometry}
\usepackage{setspace}
\usepackage{wrapfig}
\usepackage{hyperref}
\usepackage[utf8]{inputenc}
\usepackage{graphicx}
\usepackage{flushend}
\usepackage[linesnumbered,ruled]{algorithm2e}

\usepackage[font={footnotesize}]{caption}
\usepackage[font={footnotesize}]{subcaption}
\usepackage{footnote}
\usepackage{multicol}
\usepackage{multirow}
\usepackage{enumitem}   
\usepackage{soul}
\usepackage{framed}
\usepackage[titletoc,title]{appendix}
\usepackage{bm}
\usepackage{siunitx}
\usepackage{cite}
\usepackage[makeroom]{cancel} 
\usepackage{natbib}
\usepackage{hyperref}
\definecolor{darkblue}{rgb}{0,0,1}
\hypersetup{pdftex=true, colorlinks=true, breaklinks=true, linkcolor=magenta, menucolor=magenta, citecolor=magenta, urlcolor=magenta}

\newenvironment{rcases}
{\left.\begin{aligned}}
	{\end{aligned}\right\rbrace}

\definecolor{darkblue}{rgb}{0,0,1}
\hypersetup{pdftex=true, colorlinks=true, breaklinks=true, linkcolor=darkblue, menucolor=darkblue, pagecolor=darkblue, citecolor=darkblue, urlcolor=darkblue}

\usepackage[dvipsnames]{xcolor}
\newcommand{\eq}[1]{Eq.~\ref{#1}}

\newcommand{\ms}[1]{\mathrm{#1}} 
\newcommand{\mb}[1]{\mathbf{#1}} 
\newcommand{\pd}[2]{\frac{\partial #1}{\partial #2}}
\newcommand{\trr}[1]{{#1}^{\!\top}}

\begin{document}
	
	\begin{center}
		\Large{\bf{Topology Optimization and 3D-printing of Large Deformation Compliant Mechanisms for Straining Biological Tissues}}\\
		
	\end{center}
	
	\begin{center}

		\large{P. Kumar$^{\ast,}$\footnote{Corresponding author: pkumar@mek.dtu.dk,\,prabhatkumar.rns@gmail.com}, C. Schmidleithner$^\dagger$, N. B. Larsen$^\dagger$, and 	O. Sigmund$^{\ast}$}
		\vspace{4mm}
		
		\small{\textit{$^\ast$Department of Mechanical Engineering, Solid Mechanics, Technical University of Denmark,
				2800 Kgs. Lyngby, Denmark}}
			
		\small{\textit{$^\dagger$Department of Health Technology, Technical University of Denmark,
				2800 Kgs. Lyngby, Denmark}}
			
		\vspace{6mm}
	Published\footnote{This pdf is the personal version of an article whose final publication is available at \href{https://link.springer.com/article/10.1007/s00158-020-02764-4}{Structural and Multidisciplinary Optimization}}\,\,\,in \textit{Structural and Multidisciplinary Optimization}, 
	\href{https://doi.org/10.1007/s00158-020-02764-4}{DOI:10.1007/s00158-020-02764-4} \\
	Submitted on 31~March 2020, Revised on 19~August 2020, Accepted on 08~October 2020
		
	\end{center}
	
	\vspace{1mm}
	\rule{\linewidth}{.15mm}
	{\bf Abstract:}
	 This paper presents a synthesis approach in a density-based topology optimization setting to design large deformation compliant mechanisms for inducing desired strains in biological tissues. The modelling is based on geometrical nonlinearity together with a suitably chosen hypereleastic material model, wherein the mechanical equilibrium equations are solved using the total Lagrangian finite element formulation. An objective based on least-square error with respect to target strains is formulated and minimized with the given set of constraints and the appropriate surroundings of the tissues. To circumvent numerical instabilities arising due to large deformation in low stiffness design regions during topology optimization, a strain-energy based interpolation scheme is employed. The approach uses an extended robust formulation i.e. the eroded, intermediate and dilated projections for the design description as well as variation in tissue stiffness. Efficacy of the synthesis approach is demonstrated by designing various compliant mechanisms for providing different target strains in biological tissue constructs. Optimized compliant mechanisms are 3D-printed and their performances are recorded in a simplified experiment and compared with simulation results obtained by a commercial software.  \\
	

	\vspace{-4mm}
	\rule{\linewidth}{.15mm}
	
		\section{Introduction} \label{Sec:Introduction}
 Development of new drugs is challenged by the limited predictive accuracy of current simple cell models on safety and efficacy in the human body \citep{mordwinkin2013patient}. Functional mini-organ models with higher predictive value are increasingly used in the pharmaceutical industry to meet this challenge \citep{ikeda2017vitro}. These mini-organ models can be derived from  a healthy/diseased person's tissues,  adult stem cells (which can be differentiated into the particular type of tissues \textit{in vitro} relatively faster and in few steps),  human embryonic stem cells (hESCs)\footnote{Ethical issues} or by inducing pluripotency in human adult cells (hiPSCs) \citep{duelen2017stem}. To facilitate maturation of differentiated tissue cells, local static- and dynamic-mechanical forces are essential to induce the required strains \citep{vining2017mechanical}. In general, uni-axial stretching up to $15-20$\% in mini-organs (e.g., skeletal- and cardio-myocytes) is needed \citep{vandenburgh1995response} for achieving alignment and proper contractile behavior. However, current available culture systems are limited in providing the needed strains \citep{riehl2012mechanical,cook2016characterization} and, in general, are designed by trial and error approaches. 
 
 The motive herein is to provide a systematic approach, e.g., topology optimization, while considering geometric nonlinearity to design largely deformable compliant mechanisms which can furnish the required strains in the biological tissues in response to external stimuli. The nonlinearity stems from the large desired strain the biological tissues.

A compliant mechanism (CM) performs its tasks using motion obtained from the elastic deformations of its members. These mechanisms, in general, have monolithic designs and their amount of elastic deformations (small/large) depend upon the applications they are designed for. By virtue of their geometrical features, CMs offer numerous advantages over their classical rigid body counterparts, such as low manufacturing and assembly cost, less frictional losses due to absence of joints, low wear and tear, high precision and repeatability, to name a few \citep{sigmund1997design,frecker1997topological}. Thus, the use of such mechanisms designed by topology optimization is continuously increasing in a wide variety of applications e.g. path generation \citep{pedersen2001topology,saxena2001topology,kumar2019computational}, displacement deliminators \citep{saxena2013contact}, MEMS \citep{ananthasuresh1994strategies,jonsmann1999compliant}, in biomedical/~biomechanics/drug-discovery \citep{Frecker2005,kollimada2017micro,kumar2019compliant}. To add to the list of their ever expanding applications, herein, we propose an approach  using topology optimization in the nonlinear continuum regime to design compliant (micro-)mechanisms which can induce programmable strain  up to 20\% in biological tissues.

Topology optimization (TO) relocates material in an optimum fashion within a prescribed design domain by extremizing  desired objective(s) with a given set of constraints~\citep{sigmund2013topology}. In a general structural TO framework, finite elements (FEs) are employed to describe the design domain, and each FE is assigned a design (density) variable $\rho$. $\rho_i =1$ implies, $i^\mathrm{th}$ FE is in solid phase, whereas $\rho_i=0$ represents its void state. In a gradient-based TO, FEs with $0<\rho<1$ appear due to the relaxation. To discourage intermediate design values ($0<\rho<1$) and impose length scale in the final solutions, penalization and a robust formulation with Heaviside projection filter \citep{wang2011projection} is adopted in this paper.

In large deformation TO as considered here, FEs with low stiffness are prone to undergo excessive distortion/deformation and thus, cause numerical instabilities which may jeopardize the progress of optimization.  To circumvent these numerical instabilities, \cite{buhl2000stiffness} modified the Newton-Raphson convergence criterion in their approach by excluding the internal nodal forces originating from low stiffness FEs. \cite{bruns2001topology} treated such instabilities via removing and reintroducing low stiffness FEs and others suggested only to remove elements in low density regions~\citep{cho2006topology}. \cite{yoon2005element} proposed a connectivity parameterization approach by employing fictitious springs to connect FEs. \cite{lahuerta2013towards, klarbring2013topology} and \cite{luo2015topology}  employed special hyperelastic material laws for low stiffness FEs in their approaches. An approach based on scaling of the element deformation was proposed by \cite{van2014element}. \cite{wang2014interpolation} proposed a method based on an energy interpolation scheme that models low density elements as linear. Based on our extensive experience from other applications, we find that treating low density FEs as linear \citep{wang2014interpolation}, provides stable and reliable convergence and is hence used here.

In general, a bioreactor for muscle tissue maturation is expected to induce uni-axial cyclic strain up to 20\% for cellular alignment and proper contractile behavior in addition to supporting the required auxotonic resistance \citep{vandenburgh2008drug}. Numerous bioreactors for static uni-axial straining via auxotonic resistance have been reported. In those, a bioreactor with flexible poles approach  \citep{vandenburgh2008drug} requires minimum assembly, provides a method for high throughput solutions supporting the formation of 3D microsized tissue constructs, all in a low cost format. The mechanical challenge of stably attaching slippery microtissues to the flexible poles is solved by making the tissue form by self-aggregation of muscle cells seeded in a compliant gel matrix between and around the poles \citep{vandenburgh2008drug,hansen2010development}. This is a well-established procedure that was also recently demonstrated in fully 3D printed devices \citep{christensen20193d}. However, there is no medium for inducing cyclic strains mechanically in the tissue constructs. Indeed, this shortcoming of the approach motivated us to formulate the presented design problem. Compliant (micro-)mechanisms are envisioned to be combined in a design with flexible poles environment for inducing the required strains in the constructs mechanically in response to external mechanical stimuli applied on a base platform (see Sec.~\ref{Sec:SET2}). The approach is conceptualized to support many pairs of flexible poles requiring different straining in their respective biological tissues.  

Development (amount and location) of stress/strain in a structure significantly depends upon its geometrical features and the loading conditions. In TO, one can find approaches that impose stress constraints \citep{duysinx1998topology,luo2013enhanced,da2019topology}, and also, on stress isolation at predefined regions within the design domain \citep{li2014stress,luo2017optimal, picelli2018stress}. Our aim here is not to isolate strain \citep{picelli2018stress} in the design domain but rather to achieve a target strain level in the biological tissue substrates using optimized compliant mechanisms. A least square objective (see Sec.~\ref{Sec:Objectiveformulation}) based on target strains is minimized with a given set of resource constraints. Geometric  nonlinearity is considered wherein an energy interpolation scheme \citep{wang2014interpolation} is exploited to handle numerical issues arising due to large deformations in the low stiffness regions during the topology optimization. Here, the robust formulation \citep{wang2011projection} is employed in a large deformation density-based TO setting, wherein the maximum value of the three objectives, evaluated for the dilated, intermediate and eroded designs, is minimized. Further, the robust formulation is suitably modified to also accommodate varying biological tissue geometries with the flexible poles surrounding \citep{vandenburgh2008drug}. Prototypes of the designs  are printed and their performances are compared with corresponding ABAQUS analyses.  

 In summary, the contributions of this paper are: 
\let\labelitemi\labelitemii
\begin{itemize}
	\item Formulation of the topology optimization problem for designing compliant mechanisms which can induce  desired strains in biological tissues in response to external loading,
	\item Conceptualization of an objective based on least-square error which is minimized to achieve the optimum material layout for the compliant actuators,
	\item Illustration of the robust approach in a large deformation topology optimization setting for designing CMs to strain tissue constructs, which is further modified to cater for tissue construct variations,
	\item Demonstration of the approach by synthesizing various compliant mechanisms for straining biological tissue with and without flexible poles environment,
	\item Comparison of performances of the optimized mechanism designs with a commercial software, ABAQUS,
	\item Realization of the optimized mechanisms by 3D-printing and performing a simplified experiment, and also, comparing the experimental result with corresponding ABAQUS analysis result.
\end{itemize}

The layout of this paper is arranged as follows. Section~\ref{Sec:Problem Definition and Numerical Technique} describes the problem definition with optimization formulation. For the sake of completeness, we present a brief description of nonlinear finite element analysis and the energy interpolation formulation \citep{wang2014interpolation}.  The objective based on the target strains and its sensitivity analysis with energy interpolation method are presented in Section~\ref{Sec:Objectiveandsensitivity}. Section~\ref{Sec:NumericalExampleandDiscussion} presents two set of numerical examples with discussions. Optimized mechanisms with the flexible poles setting are fabricated and an experiment is performed, and their performances are compared with respective ABAQUS analyses. Lastly, in Section~\ref{Sec:Conclusions}, conclusions are drawn.

	\section{Problem Definition and Numerical Technique} \label{Sec:Problem Definition and Numerical Technique}

\begin{figure}[h!]
	\centering
	\includegraphics[scale = 1.25]{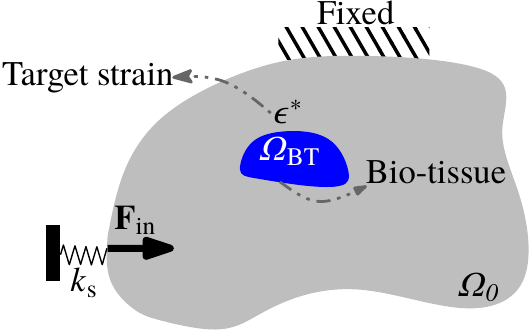}
	\caption{A schematic representation of the micro-actuator design problem. $\Omega_0$, $\Omega_{\ms{BT}}$ indicate the design domain and region for a biological tissue, respectively. $\mb{F}_\text{in}$ denotes the actuating force and $k_\text{s}$ is the input spring stiffness.}
	\label{fig:schematicfigure}
\end{figure}

A schematic diagram of the design problem  is illustrated in Fig.~\ref{fig:schematicfigure}. The aim is to obtain an optimized mechanism using TO in $\Omega_0$ which can induce the target strains $\epsilon^*$ (up to 10-20\%) in the biological tissue $\Omega_{\ms{BT}}$ in response to external actuation $\mb{F}_\text{in}$ (Fig.~\ref{fig:schematicfigure}). 
A density-based topology optimization approach \citep{sigmund200199,bendsoe2003topology} is adopted wherein each FE is assigned a design variable $\rho_e\in [0,\,1]$. The design variable is considered constant within each FE. The physical density $\bar{\rho}_e$ of an FE is defined using the smooth Heaviside projection filter \citep{wang2011projection} as
\begin{equation}\label{eq:Heaviside}
	\bar{\rho}_e(\tilde{\rho_e}(\rho_e)) = \frac{\tanh(\beta\eta) + \tanh(\beta(\tilde{\rho}_e -\eta))}{\tanh(\beta\eta) + \tanh(\beta(1-\tilde{\rho_e}))},
\end{equation}
where $\beta\in [1,\,\infty)$ defines the steepness of the projection filter and $\eta\in [0,\,1]$ is a threshold parameter for $\bar{\rho}_e$, and $ \tilde{\rho_e}$ is the filtered form of ${\rho}_e$. Ideally, $\beta\to\infty$, for a discrete  ($0-1$) solution. Practically, however to maintain smooth convergence, $\beta$ is increased from an initial value $\beta_\ms{int}=1$ to a maximum value $\beta_{\text{max}}$ using a continuation strategy (see Sec. \ref{Sec:NumericalExampleandDiscussion} for specific values). 

The filtered  variable $\tilde{\rho_e}$ is evaluated as
\begin{equation}\label{eq:densityfilter}
\tilde{\rho}_e = \frac{\displaystyle\sum_{i \in n_e}w(\bm{x})v_i\rho_i}{\displaystyle\sum_{i \in n_e} w(\bm{x})v_i},
\end{equation}
where $n_e = \{i,\,||\mb{x}_i^c - \mb{x}_e^c||\le r_{\ms{\min}}\}$ with  $\mb{x}_i^c\,\text{and}\, \mb{x}_e^c$ as center coordinates of the $i^\ms{th}$ and $e^\ms{th}$ elements respectively, $r_\ms{\min}$ is the filter radius and $||\,.\,||$ denotes distance in the Euclidean space. $v_i$ denotes the volume of $i^\ms{th}$ element, and $w(\bm{x})$, a linearly decaying weighting function, is defined as
\begin{equation}\label{eq:weightfunction}
w(\bm{x}) = r_{\mathrm{min}} - ||\mb{x}_i^c - \mb{x}_e^c||.
\end{equation}

We use the modified SIMP (Solid Isotropic Material with Penalization) interpolation scheme  to relate the physical density $\bar{\rho}_e$ with the Young's modulus of the given material as
\begin{equation}\label{eq:SIMP}
E_e(\bar{\rho_e}) = E_\ms{v} + (E_\ms{s} - E_\ms{v})(\bar{\rho_e})^{p}, \qquad \bar{\rho_e}\in[0,\,1],
\end{equation} 
where $E_\ms{s}$ and $E_\ms{v}$ are Young's moduli of the actual and void material, respectively. $E_\ms{v}$ is set to $E_\ms{s}\times 10^{-6}$ and the penalty parameter ${p}=3$ is chosen, which guides topology optimization results towards close to \textquotedblleft0-1" solutions.
\subsection{Optimization problem formulation}\label{Sec:OptimizationProblemformulation}	
 To avoid checkerboards, one-node connected hinges, mesh-dependencies, and other artifacts, \cite{sigmund2009manufacturing} formulated the design problem in a robust way wherein he employed two projection filters. \cite{wang2011projection} modified the formulation using the smooth Heaviside projection filters, which is considered herein with suitable modifications for the large deformation continuum setting. The formulation in \citep{wang2011projection} considered a set of three designs i.e. dilated, intermediate and eroded continua for a problem and minimized the worst objective obtained out of three designs. Dilate and erode are morphological-based image operators. As per \citep{sigmund2007morphology}, they can be used in a TO setting for controlling the feature sizes and ensure robustness. Designs obtained using these operation in association with suitable filtering are called dilated and eroded structures. The dilate operation corresponds to under-etching of the fabricated designs, whereas erode operation corresponds to over-etching. The intermediate designs indicate the desired or correctly etched structures, i.e., the blueprint structures.

The dilated $\bar{\bm{\rho}}^d$, intermediate $\bar{\bm{\rho}}^i$ and eroded $\bar{\bm{\rho}}^e$ design vectors are obtained via \eq{eq:Heaviside} using the threshold $0.5-\Delta\eta,\,0.5$ and $0.5+\Delta\eta$, respectively (see Sec.~\ref{Sec:NumericalExampleandDiscussion} for specific $\Delta\eta$). The optimization problem is formulated in the nonlinear continuum setting as a min/max problem \citep{wang2011projection} to also accommodate different geometries of the tissue construct, which can be written as: 
\begin{equation}\label{eq:actualoptimization}
	\small
	\begin{rcases}
	& \underset{\bm{\rho}}{\text{min}}:\underset{k}{\text{max}}
	& &: \left(f_{k}(\mb{u}_k,\,\bar{\bm{\rho}}^d(\bm{\rho})),\,f_{k}(\mb{u}_k,\,\bar{\bm{\rho}}^i(\bm{\rho})),\,f_{k}(\mb{u}_k,\,\bar{\bm{\rho}}^e(\bm{\rho}))\right) \\
	& \textit{s.t.} & &:\mb{R}_k(\mb{u}_k,\bar{\bm{\rho}}^d(\bm{\rho})) = \mb{0}, k=1,\,2,\cdots,N_\text{BT} \\
	& & &:\mb{R}_k(\mb{u}_k,\bar{\bm{\rho}}^i(\bm{\rho})) = \mb{0}, k=1,\,2,\cdots,N_\text{BT} \\
	& & &:\mb{R}_k(\mb{u}_k,\bar{\bm{\rho}}^e(\bm{\rho})) = \mb{0}, k=1,\,2,\cdots,N_\text{BT} \\
	&  &&: V_f(\bar{\bm{\rho}}^d(\bm{\rho}))-V_d^*\le 0 \\
	& &&:\mb{0}\le\bm{\rho}\le\mb{1}\\
	\end{rcases},
	\end{equation}
where $\mb{R}_k(\mb{u}_k,\bar{\bm{\rho}}^d(\bm{\rho}))$, $\mb{R}_k(\mb{u}_k,\bar{\bm{\rho}}^i(\bm{\rho}))$ and $\mb{R}_k(\mb{u}_k,\bar{\bm{\rho}}^e(\bm{\rho}))$ are the residual terms (Eq.~\ref{eq:residualform}) for the dilated, intermediate and eroded designs, respectively, $f_k$ is the formulated objective (see Sec.~\ref{Sec:Objectiveformulation}) and $N_\text{BT}$ is the number of variations of the tissue construct. $V_f(\bar{\bm{\rho}}^d(\bm{\rho}))$ and $V_d^*$ are the volume fraction and its upper limit for the dilated design, respectively. Volume of the dilated design  is updated after every specific number of optimization iteration so that the volume fraction of the intermediate design becomes equal to the prescribed volume $V^*_i$ at the end of the optimization process when the volume constraint is active \citep{wang2011projection}. The formulation furnishes three material distributions and $3\times N_\text{BT}$ load cases but only one design variable field for a design, one can select as per the manufacturing and material limits.  However, the intermediate design is for the intended blueprint realization.

\subsection{Finite element formulation} \label{Sec:Finiteelementformulation}
 For the sake of completeness, we present the used nonlinear finite element approach \citep{zienkiewicz2005finite,bathe2006finite,wriggers2008nonlinear} in brief here. The total Lagrangian finite element formulation is considered in this approach, and all desired strains refer to the original coordinate system.
The deformation gradient $\bm{F}$ is defined\footnote{Italic font is used to write the field quantities, whereas the discrete quantities are written using normal font.} as
\begin{equation}\label{eq:deformationgradient}
\bm{F} = \bm{I} + \nabla_0\bm{u},
\end{equation}
where $\bm{I}$ is the identity tensor and $\nabla_0\bm{u}$ indicates gradient of the displacement field with respect to reference configuration $\bm{X}\in\Omega_0$. In terms of $\bm{F}$, the right Cauchy-Green deformation tensor $\bm{C}$ equals to $\trr{\bm{F}}\bm{F}$, which is used further to find the Green-Lagrangian strain tensor $\bm{E}$ as
\begin{equation}\label{eq:straindefinition}
\bm{E} = \frac{1}{2} (\bm{C} -\bm{I}).
\end{equation}

In view of the standard FE method, the weak form of a mechanical equilibrium equation provides \citep{zienkiewicz2005finite}
\begin{equation}\label{eq:residualform}
\mb{R}(\mb{u},\bm{\rho}) = \mb{F}^\ms{int}(\mb{u},\bm{\rho}) - \mb{F}^\ms{ext}= \mb{0},
\end{equation}
where $\mb{R}(\mb{u},\bm{\rho})$ is a residual term. The internal force at element level $\mb{F}_e^\ms{int}$ is evaluated as
\begin{equation}\label{eq:internalforce}
\mb{F}_e^\ms{int} = \int_{\Omega_0^e}\trr{\mb{B}}_\ms{TL}(\mb{u})\mb{S}_e(\mb{u},\,\bm{\rho})\,d \Omega_0^e,
\end{equation}
  where $\mb{B}_\ms{TL}(\mb{u})$ and $\mb{S}_e$  are the total Lagrangian strain-displacement matrix \citep{bathe2006finite} and the second Piola-Kirchhoff stress tensor of an FE $\Omega_0^e$, respectively.  $\mb{F}^\ms{ext}$ is assumed to be a constant force here. \eq{eq:residualform} is solved herein using the Newton-Raphson (N-R) iterative solver. 

The second Piola-Kirchhoff stress tensor is evaluated as $\bm{S} = 2\frac{\partial W}{\partial \bm{C}}$ using strain-energy function W and tensor $\bm{C}$. W is taken here as \citep{zienkiewicz2005finite}
\begin{equation}\label{eq:strainenergyfunction}
W = \frac{G}{2}\left(\mathcal{J}_{1\bm{C}} -3-\ln J\right) + \frac{\kappa}{2}(J -1)^2,
\end{equation}
where $G = \frac{E}{2(1+\nu)}$  is the shear modulus, however one evaluates $\kappa = \frac{E}{2(1-\nu)}$ and $\kappa = \frac{E}{2(1+\nu)(1-2\nu)}$ for 2D plane stress and 2D plane strain, respectively. $\mathcal{J}_{1\bm{C}}$ is the first principal invariant of the right Cauchy-Green tensor $\bm{C}$ and $J=\det \bm{F}$. Further, $E$ denotes Young's modulus and $\nu$ indicates the Poisson's ratio. Note that the employed material model (Eq.~\ref{eq:strainenergyfunction}) accounts for both geometric and material nonlinearities of the tissues. Using the fundamentals of nonlinear continuum mechanics \citep{holzapfel2001nonlinear}, one finds the second Piola Kirchhoff stress $\bm{S} = G(\bm{I} - \bm{C}^{-1}) + \kappa (J -1)J\bm{C}^{-1}$ for the given strain-energy function (\eq{eq:strainenergyfunction}).

\subsection{Numerical treatment of low stiffness regions}\label{Sec:NumericalTreatment}

 To handle numerical instabilities in the TO due to large deformation, the energy interpolation scheme \citep{wang2014interpolation} which ensures smooth deformation, is employed herein. Mathematically, the interpolated strain-energy $\bar{W_e}$ for an FE is written as

\begin{equation}
\bar{W_e}(\mb{u}_e) = \left[W_e(\gamma_e\bm{u}_e) - W_e^L(\gamma_e\bm{u}_e) + W_e^L(\bm{u}_e)\right] E_e (\rho_e), 
\end{equation}  
where $W_e (.)$ indicates the strain-energy density function of the actual material at unit Young's modulus, $W_e^L(.)$ is the strain-energy for small deformation at unit Young's modulus, and $E_e$ is the Young's modulus of an element (\eq{eq:SIMP}). Further, $\gamma_e$, defined using a smooth Heaviside projection filter, is used to determine behavior of a FE as
\begin{equation}\label{eq:gammadefinition}
\gamma_e = \frac{\tanh(\beta_1\eta_0) + \tanh(\beta_1({\bar{\rho}_e}^\text{p} -\eta_0))}{\tanh(\beta_1\eta_0) + \tanh(\beta_1(1-\eta_0))},
\end{equation} 
where $\eta_0$ is the threshold value. We use $\beta_1 = 500$ and $\eta_0=0.01$ for all examples as suggested by \citep{wang2014interpolation}.

\section{Formulation of Objective Function and Sensitivity Analysis}\label{Sec:Objectiveandsensitivity}
This section describes formulation of the objective and its sensitivity analysis.
\subsection{Formulation of objective function}\label{Sec:Objectiveformulation}
 
 Let $\epsilon_\ms{xx}^*$, $\epsilon_\ms{yy}^*$ and $\epsilon_\ms{xy}^*$ indicate the target strains in $x-$, $y-$ and $xy \,(\text{shear})-$directions respectively. Then, the error objective $f_k$ can be formulated in a continuum setting as

\begin{equation}\label{eq:objective}
\begin{aligned}
f_k = \frac{1}{A}\int_{\Omega_{\ms{BT}}}\left(\frac{w_1(^k\epsilon_\ms{xx}^e -\epsilon_\ms{xx}^*)^2 + w_2 (^k\epsilon_\ms{yy}^e -\epsilon_\ms{yy}^*)^2+ w_3(^k\epsilon_\ms{xy}^e -\epsilon_\ms{xy}^*)^2 }{w_1(\epsilon_\ms{xx}^*)^2 + w_2(\epsilon_\ms{yy}^*)^2 + w_3(\epsilon_\ms{xy}^*)^2}\right)dA,
\end{aligned}
\end{equation}
\noindent and in its corresponding FE setting evaluated in element centroids
\begin{equation}\label{eq:objective_FE}
\begin{aligned}
f_k = \frac{1}{N_{be}}\displaystyle\sum_{i= 1}^{N_{be}}\left(\frac{w_1(^k\epsilon_\ms{xx}^i -\epsilon_\ms{xx}^*)^2 + w_2 (^k\epsilon_\ms{yy}^i -\epsilon_\ms{yy}^*)^2+ w_3(^k\epsilon_\ms{xy}^i -\epsilon_\ms{xy}^*)^2 }{w_1(\epsilon_\ms{xx}^*)^2 + w_2(\epsilon_\ms{yy}^*)^2 + w_3(\epsilon_\ms{xy}^*)^2}\right),
\end{aligned}
\end{equation}
where $w_1,\,w_2,\,\text{and}\, w_3$ are user defined weighing factors depending upon the desired axis of straining.  $\Omega_{\ms{BT}}$ is the design region for the biological tissue and $A$ is the associated area. Further, $N_{be}$ is the total number of FEs used to represent $\Omega_{\ms{BT}}$. In the Voigt notation\footnote{Employed here to represent the stress, strain and material tangent tensors for the FE analysis.} for a 2D case,  $^k\epsilon_\ms{xx}^i$, $^k\epsilon_\ms{yy}^i$ and $^k\epsilon_\ms{xy}^i$ are the first, second and third entries of the Green-Lagrange strain $^k\mb{E}_e$ which is evaluated at the center, of the $i^\ms{th}$ FE associated to the $k^\text{th}$ biological tissue. Note that strains are not very accurately modelled at the center points when using the standard finite element approaches. However, we are minimizing a function which is an integral over a large area and hence, errors are expected to be very small.

\subsection{Sensitivity analysis}\label{Sec:Sensitivityanalysis}

A  gradient-based approach is employed to solve the optimization problem (\eq{eq:actualoptimization}). Sensitivities of  the objective and the constraints with respect to the design vector $\bm{\rho}$ are evaluated using the adjoint-variable method. The augmented performance function $\mathcal{L}$, defined using the objective and the equilibrium equation (Eq.~\ref{eq:residualform}), can be written as
\begin{equation}\label{eq:augmented objective}
\mathcal{L}  = f_k(\mb{u}_k,\bm{\bar{\rho}}) + \trr{\bm{\lambda}}\mb{R}_k(\mb{u}_k,\bm{\bar{\rho}}),
\end{equation}   
where $\bm{\lambda}$ is the Lagrange multiplier vector. Differentiating 
\eq{eq:augmented objective} with respect to $\bar{\bm{\rho}}$ yields
\begin{equation}\label{eq:lagderivative}
\frac{d \mathcal{L}}{d \bar{\bm{\rho}}} = \pd{f_k}{\bar{\bm{\rho}}} + \underbrace{\left[\pd{f_k}{\mb{u}_k} + \trr{\bm{\lambda}}\pd{\mb{R}_k}{\mb{u}_k}\right]}_{\text{Term 1}}\pd{\mb{u}_k}{\bar{\bm{\rho}}} + \trr{\bm{\lambda}}\pd{\mb{R}_k}{\bar{\bm{\rho}}}.
\end{equation}
One chooses $\bm{\lambda}$ such that Term 1 vanishes \citep{bendsoe2003topology}, i.e.,
\begin{equation}\label{eq:langarangevalue}
\trr{\bm{\lambda}}\mb{K}_\text{T} = - \pd{f_k}{\mb{u}_k},
\end{equation}
where $\mb{K}_\text{T} = \pd{\mb{R}_k}{\mb{u}_k}$ \citep{zienkiewicz2005finite} is the tangent stiffness matrix at the equilibrium state \citep{buhl2000stiffness}. A procedure to evaluate $\pd{f_k}{\mb{u}_k}$ is mentioned in Appendix \ref{Append1}. $\pd{\mb{R}_k}{\bar{\bm{\rho}}}$ can be evaluated in view of \eq{eq:residualform} as 
\begin{equation} \label{eq:residualderivative}
\pd{\mb{R}_k}{\bar{\bm{\rho}}} = \left(\pd{\mb{F}^\ms{int}}{\bar{\bm{\rho}}} + \pd{\mb{F}^\ms{int}}{\bm{\gamma}}\pd{\bm{\gamma}}{\bar{\bm{\rho}}}\right).
\end{equation}
One can use the chain rule to evaluate the derivative $\pd{\mathcal{L}}{\bm{\rho}}$ as \citep{wang2014interpolation}
\begin{equation}
\pd{\mathcal{L}}{\bm{\rho}} = \pd{\mathcal{L}}{\bar{\bm{\rho}}}\pd{\bar{\bm{\rho}}}{\tilde{\bm{\rho}}}\pd{\tilde{\bm{\rho}}}{\bm{\rho}}.
\end{equation}
One finds $\pd{\bar{\bm{\rho}}}{\tilde{\bm{\rho}}}$ and  $\pd{\tilde{\bm{\rho}}}{\bm{\rho}}$ using Eq.~\ref{eq:Heaviside} and  Eq.~\ref{eq:densityfilter}, respectively. The method of moving asymptotes (MMA) is used \citep{svanberg1987method} to update the design vector.

\section{Numerical Examples and Discussion}\label{Sec:NumericalExampleandDiscussion}
\begin{figure}[!h]
	\centering
	\includegraphics[scale = 1]{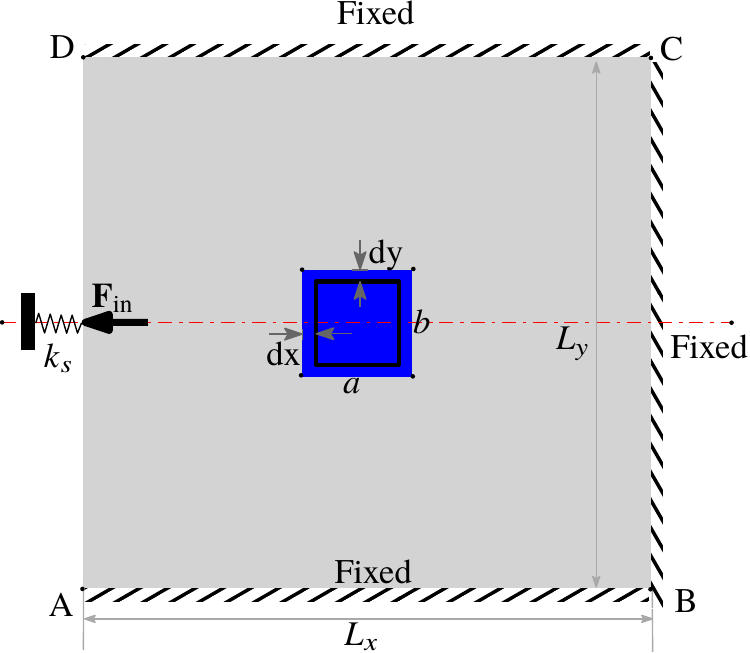}
	\caption{Design domain with a central biological tissue (blue in color). $L_x\times L_y = \SI{10}{mm}\times\SI{10}{mm}$ and  $a=\SI{2}{mm}$ and $b=\SI{2}{mm}$. $k_s$ is the stiffness of the input spring and $\mb{F}_\text{in}$ is the actuating force. The bottom, right and top sides of the domain are fixed.}
	\label{fig:AcademicDesigndomain}
\end{figure}
\begin{figure*}[!h]
	\begin{subfigure}[t]{0.45\textwidth}
		\centering
		\includegraphics[width=\linewidth]{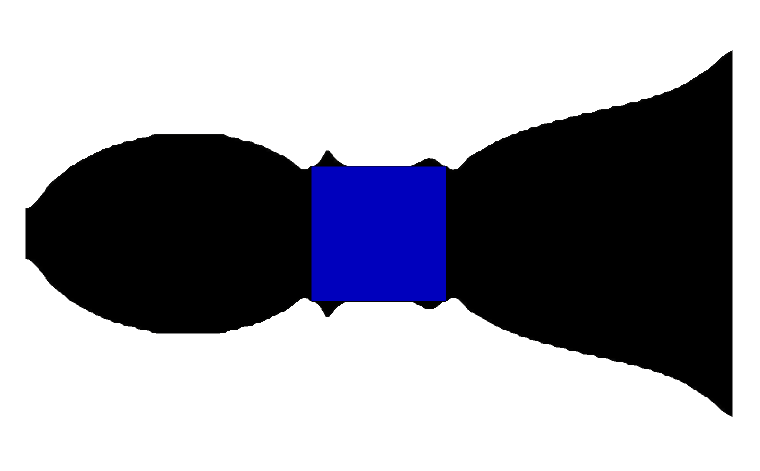}
		\caption{Intermediate design}
		\label{SET1EX1i}
	\end{subfigure}
	\begin{subfigure}[t]{0.45\textwidth}
		\centering
		\includegraphics[width=\linewidth]{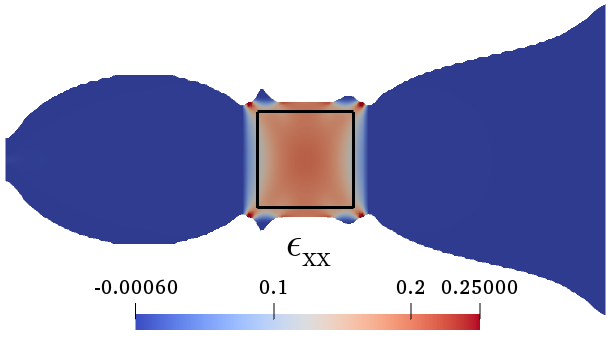}
		\caption{Actual strain distribution}
		\label{SET1EX1strain}
	\end{subfigure}
	\begin{subfigure}[t]{0.445\textwidth}
		\centering
		\includegraphics[width=\linewidth]{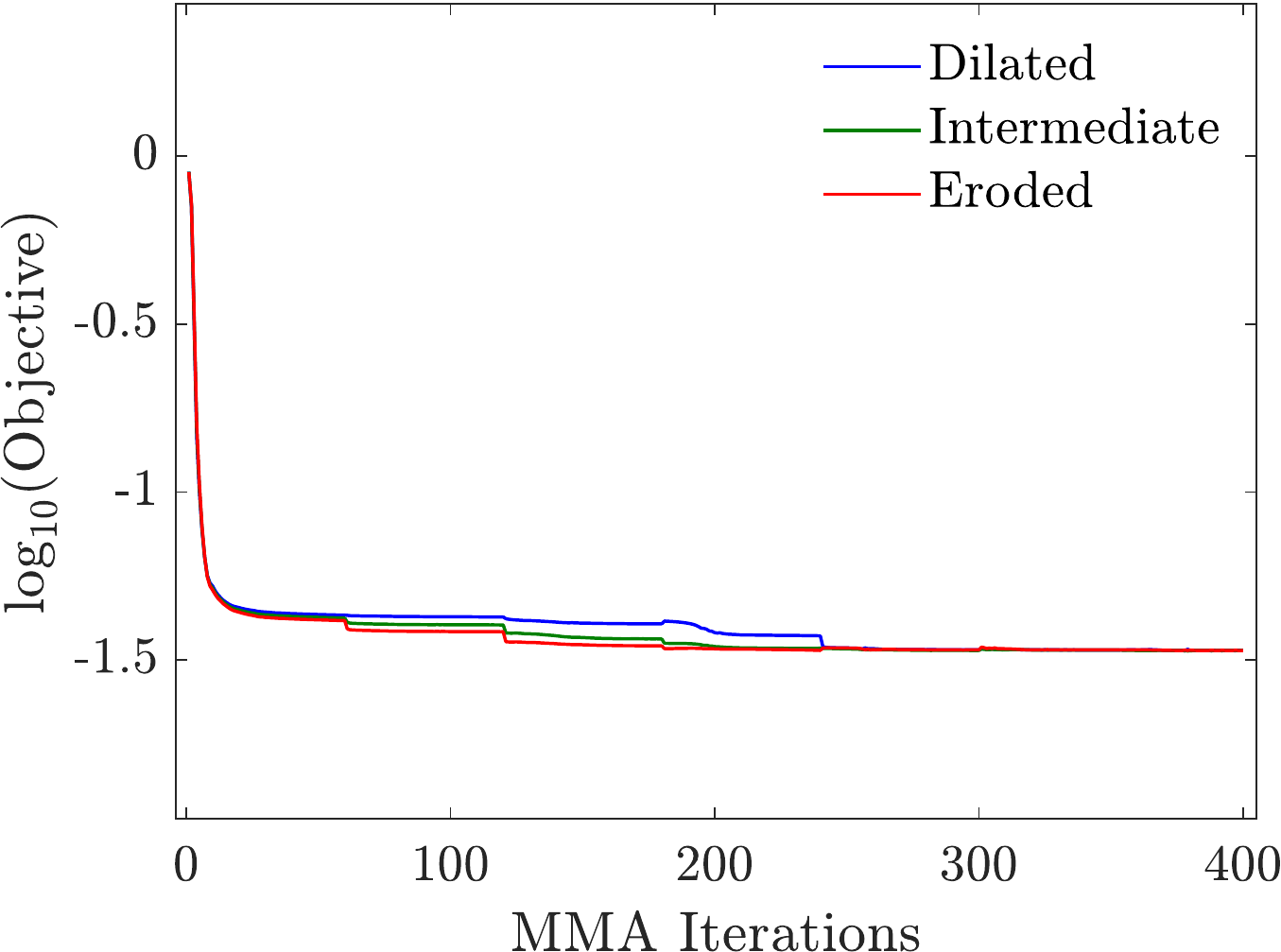}
		\caption{Objectives convergence plot}
		\label{SET1EX1ObjectiveCov}
	\end{subfigure}
	\quad
	\begin{subfigure}[t]{0.445\textwidth}
		\centering
		\includegraphics[width=\linewidth]{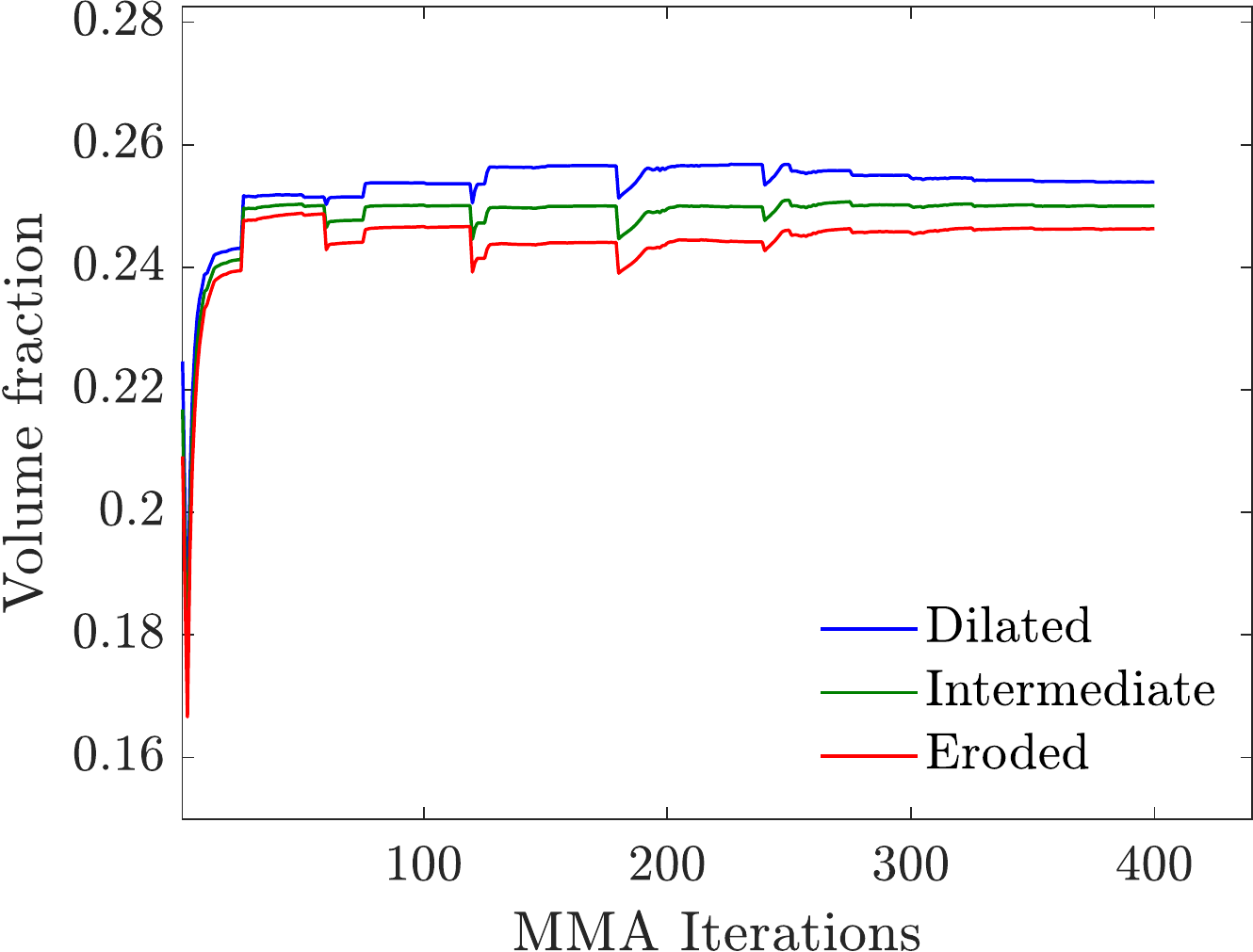}
		\caption{Volume fraction convergence plot}
		\label{SET1EX1VolumeCov}
	\end{subfigure}
	\caption{Solutions for inducing 20\% strain in $x-$direction in the central tissue. (\subref{SET1EX1i}) Optimized intermediate design, $M_\text{nd}=0.47\%$, (\subref{SET1EX1strain}) Actual axial strain distribution plot. The obtained strain distribution for the tissue within the black edged rectangle is close to uniform and approximately equal to 0.20. (\subref{SET1EX1ObjectiveCov}) Objectives convergence plot and (\subref{SET1EX1VolumeCov}) Volume fraction convergence plot. }\label{fig:SET1EX1}
\end{figure*}

This section presents two sets of numerical examples to demonstrate efficacy and robustness of the compliant micro-actuator design optimization approach. In the first set (SET-1), conceptualized academic examples with $N_\text{BT} =1$ (Eq.~\ref{eq:actualoptimization}) are solved to show controllability of strains in different directions, whereas the second set (SET-2) with $N_\text{BT} =2$ pertains to a practical application, i.e., in conjunction with the flexible poles approach \citep{vandenburgh2008drug}. In addition, the numerical results are appraised with discussions, 3D-printed prototypes, an experimental setup and results, as well as ABAQUS analyses results. 

For the examples presented in both sets, some parameters are set common as follows. The dilated, intermediate and eroded designs are evaluated using $\Delta\eta = 0.05$. External move limit, i.e., change in design variables per MMA iteration, is set to 0.1. The material definition given in Eq.~\ref{eq:strainenergyfunction} is used with Poisson's ratio $\nu=0.45$ and plane strain conditions. The volume fraction for the intermediate design is set to $V_i^*=0.25$ and the volume of the dilated design is updated every $25^\text{th}$ MMA iteration. 

\subsection{SET-1: Numerical examples}\label{Sec:SET-1: Numerical Example}
In this section, the mechanisms are designed in a general setting wherein a biological tissue (blue in color) is assumed to be placed in the middle of the design domain (Fig.~\ref{fig:AcademicDesigndomain}). 

Figure~\ref{fig:AcademicDesigndomain} indicates the design domain specifications. Length and width of the design domain ABCD are $L_x = \SI{10}{mm}$ and $L_y= \SI{10}{mm}$, respectively. The central blue domain, length $a=\SI{2}{mm}$ and width $b=\SI{2}{mm}$, denotes a biological tissue. The symmetric\footnote{Symmetric about a horizontal line} half of the design domain is parameterized using  $N_\text{ex}\times N_\text{ey} =200\times100$ quad-FEs, where $N_\text{ex}$ and $N_\text{ey}$ denote number of the FEs in $x-$ and $y-$directions respectively. The bottom, right and top sides of the domain are fixed, whereas the center of the left side of the domain is used to apply a strain-based actuator controlled by input spring with stiffness $k_s= \SI{10000}{N/m}$ and blocking force $\mb{F}_\text{in}$, as depicted in Fig.~\ref{fig:AcademicDesigndomain}.  The out-of-plane thickness for the mechanism domain and biological tissue is set to $\SI{2}{mm}$. Young's moduli for the biological tissue and remaining region (mechanism domain) are taken as $\SI{0.1}{MPa}$ and $\SI{25}{MPa}$, respectively.
\begin{figure*}[!h]
	\begin{subfigure}[t]{0.225\textwidth}
		\centering
		\includegraphics[width=0.95\linewidth]{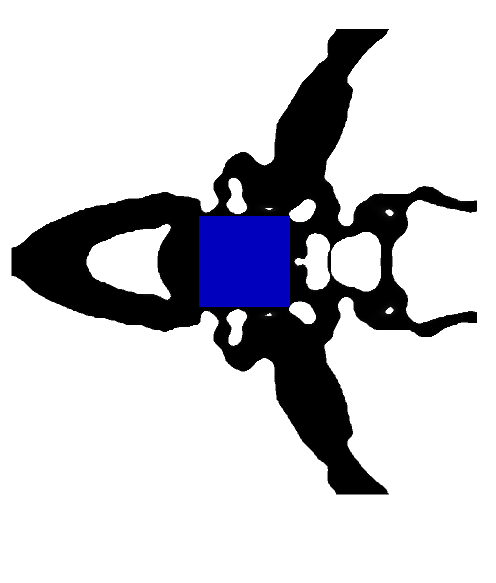}
		\caption{Intermediate design}
		\label{SET1EX2}
	\end{subfigure}
	\begin{subfigure}[t]{0.225\textwidth}
		\centering
		\includegraphics[width=\linewidth]{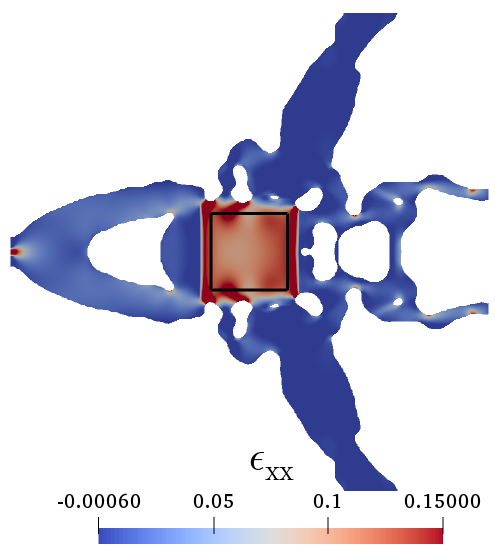}
		\caption{Strain in $x-$direction}
		\label{SET1EX2_11}
	\end{subfigure}
	\begin{subfigure}[t]{0.225\textwidth}
		\centering
		\includegraphics[width=\linewidth]{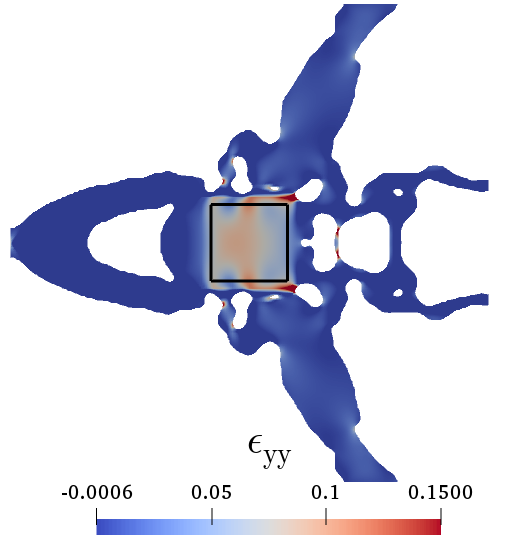}
		\caption{Strain in $y-$direction}
		\label{SET1EX2_22}
	\end{subfigure}
	\begin{subfigure}[t]{0.225\textwidth}
		\centering
		\includegraphics[width=\linewidth]{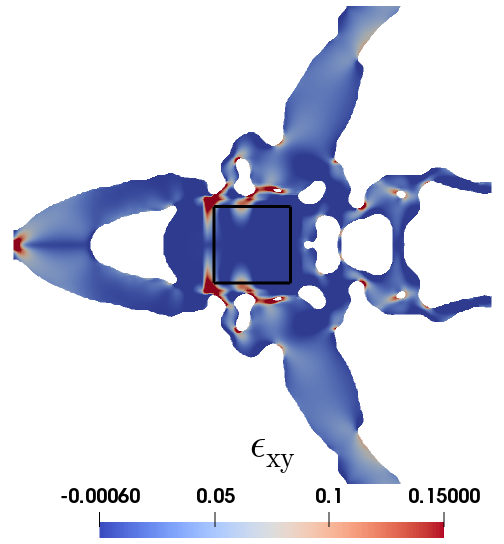}
		\caption{Shear strain}
		\label{SET1EX2_12}
	\end{subfigure}
	\caption{Solutions for inducing 12.5\%, 7.5\% and 0.0\% strains in $x-$, $y-$ and shear directions in the biological tissue. (\subref{SET1EX2}) Optimized intermediate design, $M_\text{nd}=2.91\%$ (\subref{SET1EX2_11}) Strain distribution in $x-$direction, (\subref{SET1EX2_22}) Strain distribution in $y-$direction. The maximum and minimum observed strains in $y-$direction are 0.11 and 0.035, respectively. (\subref{SET1EX2_12}) Shear strain distribution.}\label{fig:SET1EX2}
\end{figure*}

The Heaviside parameter $\beta$ is changed from $1$ to $\beta_{\text{max}}=32$ using a continuation scheme wherein it is doubled  each $60^\text{th}$ MMA iteration and once it reaches maximum value $\beta_{\text{max}}$, it remains so for the remaining optimization iterations. The filter radius is set to $5.6\times\max(\frac{L_x}{N_\text{ex}}, \frac{L_y}{2N_\text{ey}})$.   Maximum number of optimization iteration is set to 400. The objective is evaluated  within the black edged rectangle (Fig.~\ref{fig:AcademicDesigndomain}), where dx$=3\frac{L_x}{N_\text{ex}}$ and dy $=3\frac{L_y}{2N_\text{ey}}$ are taken. This is done to avoid inclusion of high localized strains, appearing on the corners of the biological tissue, in the objectives evaluation. 

The binary nature of the optimized mechanisms is measured by a gray scale indicator $M_\text{nd}$ which is defined~as~ \citep{sigmund2007morphology}
\begin{equation}\label{eq:Binarynature}
M_\text{nd} = \frac{\sum_{i=1}^{N_e}4\rho_i(1-\rho_i)}{N_e}\times 100\%.
\end{equation}
 where $N_e$ is the total number of FEs used to describe the design domain. The root mean square (RMS) errors in actual strain can be evaluated as
\begin{equation}\label{eq:errormeasure}
\begin{aligned}
\text{Err}_x = \sqrt{f_k|_{w_1 = 1,\,w_2=w_3=0}}\times 100\%,\\
\text{Err}_y = \sqrt{f_k|_{w_2 = 1,\,w_1=w_3=0}}\times 100\%,\\
\text{Err}_{xy} = \sqrt{f_k|_{w_3 = 1,\,w_1=w_2=0}}\times 100\%,\\
\end{aligned}
\end{equation}
where $\text{Err}_x$, $\text{Err}_y$ and $\text{Err}_{xy}$ are the RMS errors in the $x-$, $y-$ and shear directions, respectively.

\subsubsection{Example 1}
In this example, we seek a mechanism which can induce 20\% axial (in the $x-$direction) strain in the biological tissue (Fig.~\ref{fig:AcademicDesigndomain}) when it is actuated by a force F$_\text{in}= -\SI{3.0}{N}$ in the $x-$direction\footnote{Corresponding to an unloaded actuator with displacement $\SI{0.3}{mm}$}. $\epsilon_\ms{xx}^* = 0.20, w_1 =1$ and $w_2=w_3=0$ are used to evaluate the objectives.  

Figure~\ref{SET1EX1i} indicates the full\footnote{Suitably transferred from the symmetric half results} optimized mechanism with the central biological tissue for the intermediate design. The gray scale indicator $M_\text{nd}$ for the optimized dilated, intermediate and eroded mechanisms are evaluated to be 0.56\%, 0.47\% and 0.46\%, respectively. 

The strain distribution for $x-$direction is depicted in Fig.~\ref{SET1EX1strain}. One notices, the strain distribution in the tissue (within the black edged rectangle) is close to uniform with value approximately equal to $0.2$ which is the desired strain. However, strain near edges of the biological tissue are either lower or higher than $20\%$ (Fig. \ref{SET1EX1strain}). The $\text{Err}_x$ error in the strain is calculated using  Eq.~\ref{eq:errormeasure}, which is equal to 18.38\%. Figure~\ref{SET1EX1ObjectiveCov} and Fig.~\ref{SET1EX1VolumeCov} illustrate the objective and volume fraction convergence plots. At the end of optimization the volume constraint, 25\% volume fraction of the intermediate design, is active. A smooth convergence for both the plots can be noted at the end of the optimization.

\begin{figure*}[!h]
	\begin{subfigure}[t]{0.225\textwidth}
		\centering
		\includegraphics[width=0.95\linewidth]{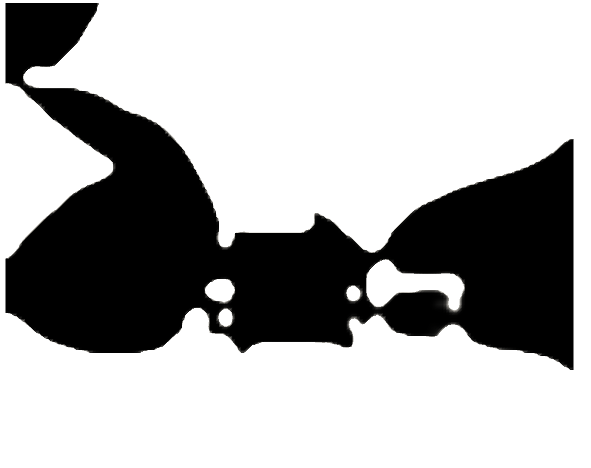}
		\caption{Intermediate design}
		\label{SET1EX3}
	\end{subfigure}
	\begin{subfigure}[t]{0.225\textwidth}
		\centering
		\includegraphics[width=\linewidth]{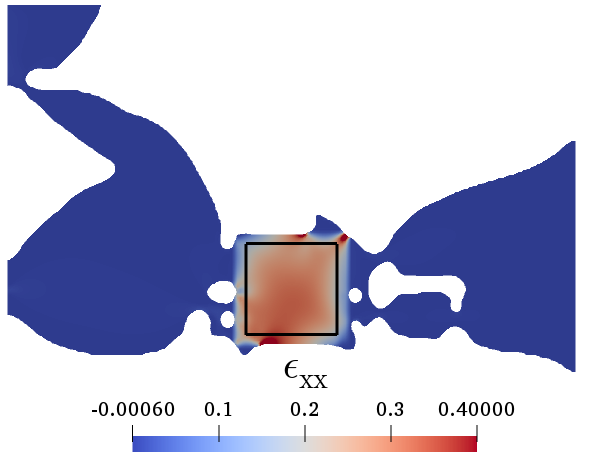}
		\caption{Strain in $x-$direction}
		\label{SET1EX3_11}
	\end{subfigure}
	\begin{subfigure}[t]{0.225\textwidth}
		\centering
		\includegraphics[width=\linewidth]{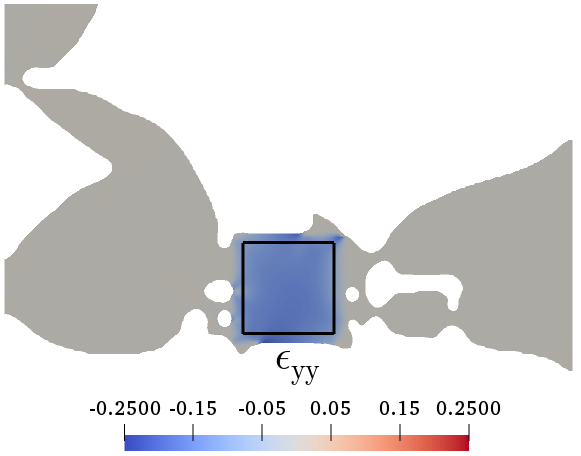}
		\caption{Strain in $y-$direction}
		\label{SET1EX3_22}
	\end{subfigure}
	\begin{subfigure}[t]{0.225\textwidth}
		\centering
		\includegraphics[width=\linewidth]{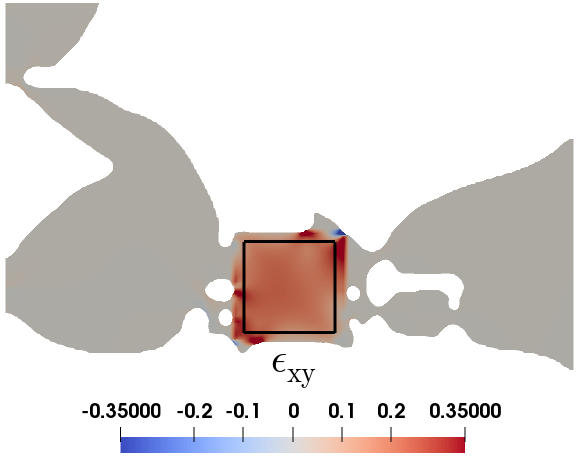}
		\caption{Shear strain}
		\label{SET1EX3_12}
	\end{subfigure}
	\caption{Solutions for inducing 30.0\%, -15.0\% and 20.0\% strains in $x-$, $y-$ and shear directions in the biological tissue. (\subref{SET1EX3}) Optimized intermediate design, $M_\text{nd}=1.13\%$ (\subref{SET1EX3_11}) Strain distribution in $x-$direction, (\subref{SET1EX3_22}) Strain distribution in $y-$direction. (\subref{SET1EX3_12}) Shear strain distribution.}\label{fig:SET1EX3}
\end{figure*}
\subsubsection{Example 2}
In this example, a mechanism which can provide a desired bi-axial straining in the biological tissue, is designed. The design domain and optimization specifications for this example are same as those used for \textit{Example 1}. The target strains $\epsilon_\ms{xx}^*=0.125$ and $\epsilon_\ms{yy}^* =0.075$ are set. The actuating force for this case is F$_\text{in}=\SI{-8.0}{N}$. The objective (Eq.~\ref{eq:objective_FE}) is evaluated for $\epsilon_\ms{xx}^*=0.125$, $\epsilon_\ms{yy}^* =0.075$, $\epsilon_\ms{xy}^*=0.0$ and $w_1=w_2=w_3=1$.

Figure~\ref{fig:SET1EX2} indicates the optimized intermediate design of the mechanism which can provide bi-axial strains in the central tissue. The actual strain distribution for the $x-$ and $y-$axes are reported in Fig.~\ref{SET1EX2_11} and Fig.~\ref{SET1EX2_22} respectively. The recorded errors in strain distribution for $x-$, $y-$ and $xy-$ (shear) directions are Err$_x = 11.87$\%, Err$_y= 11.18$\% and Err$_{xy}= 10.27\%$, respectively. One can use a higher $p-$norm, if these errors are critical. One notices that the strains at the edges of the biological tissue are higher than the desired ones (Fig.~\ref{SET1EX2_11} and Fig.~\ref{SET1EX2_22}). However, within the black-edged rectangle, the strain-distributions (Fig.~\ref{SET1EX2_11}, Fig.~\ref{SET1EX2_22} and Fig.~\ref{SET1EX2_12}), by and large, are very close to their respective target strains. $M_\text{nd} = 2.91$\% is obtained for the intermediate design (Fig. \ref{SET1EX2}). 
\begin{figure}[h!]
	\centering
	\includegraphics[width=\linewidth]{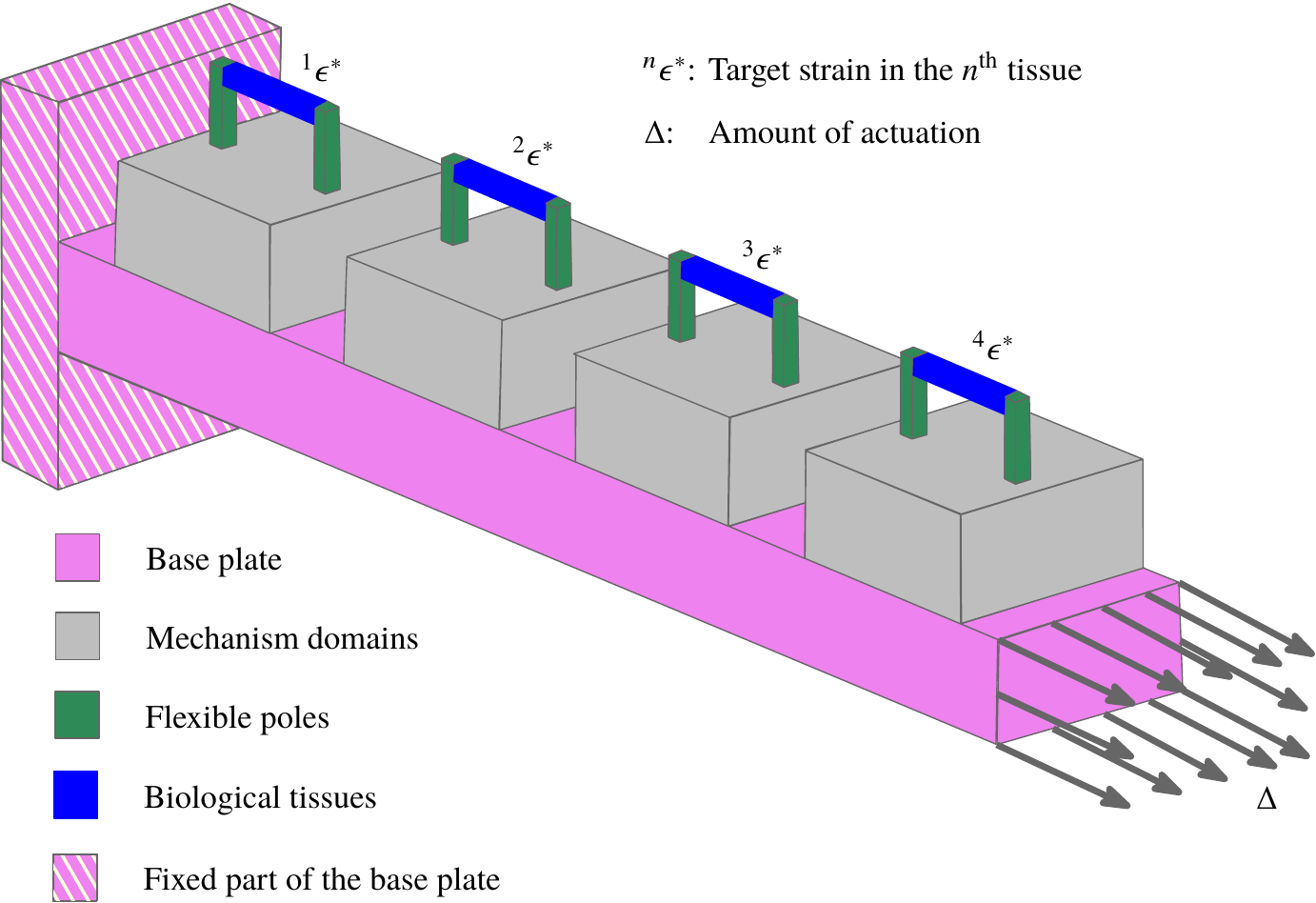}
	\caption{A schematic diagram for the design problem in a region}
	\label{fig:3DPic}
\end{figure}
\begin{figure}[h!]
	\centering
	\includegraphics[scale = 1]{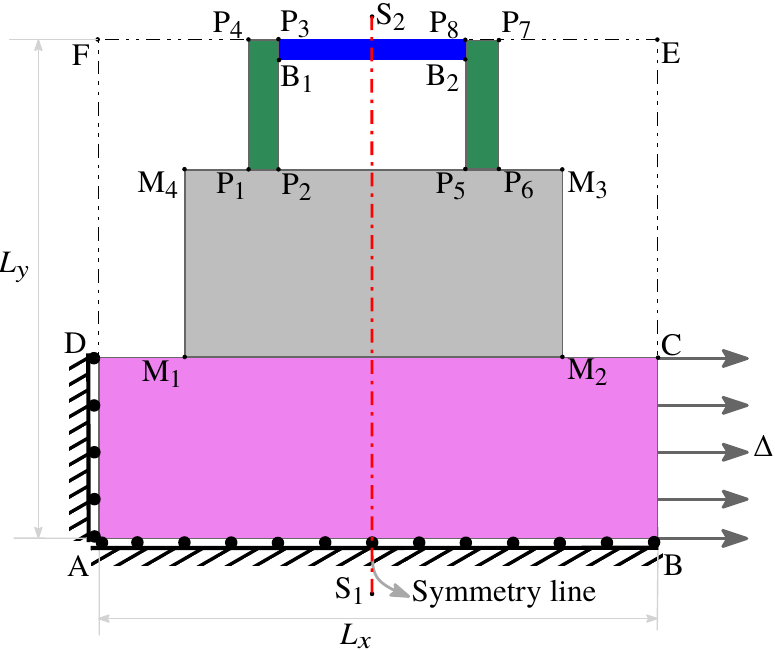}
	\caption{Design domain for designing a compliant mechanisms in conjunction with the flexible poles approach \citep{vandenburgh2008drug}. The left and bottom sides of the base plate are with roller supports. The right side of the base plate is used for applying actuation of amount $\Delta = 0.1L_x$. Figure has same color scheme as Fig.~\ref{fig:3DPic}.}
	\label{fig:ActualDesignDomain}
\end{figure}
\subsubsection{ Example 3}
This example considers inducing strains in axial, transverse and shear directions in the central biological tissue (Fig.~\ref{fig:AcademicDesigndomain}). The design domain specifications and optimization parameters are the same as those used for \textit{Example~1}. The full design domain is considered for the optimization instead of the symmetric half in order to accommodate the prescribed non-zero shear strain. The target strains are set as $\epsilon_\text{xx}^*= 0.30$, $\epsilon_\text{yy}^*=-0.15$, and $\epsilon_\text{xy}^*=0.20$  and the objective is evaluated with $w_1= w_2=w_3=1.$ The actuating force in the $x-$direction is set to F$_\text{in} = \SI{-4.5}{\newton}$. We use $200\times 200$ bi-linear FEs to describe the design domain

The optimized intermediate design of the mechanism is displayed in Fig.~\ref{SET1EX3}, which induces the desired strains  in axial (Fig.~\ref{SET1EX3_11}), transverse (Fig.~\ref{SET1EX3_22}) and shear (Fig.~\ref{SET1EX3_12}) directions in the tissue. The gray scale indicator $M_\text{nd} $ is $1.13\%$. The recorded error in axial, transverse and shear directions are Err$_x = 19.67$\%, Err$_y= 13.30$\% and Err$_{xy}= 17.74$\%, respectively. One notices that, by and large, the actual strain distributions are uniform in each direction within the black-edged rectangle and also, close to their respective desired ones.

\begin{table*}[h!]
	\centering
	\begin{tabular}{c|c|c|c|c|c}
		\textbf{Region}      & \textbf{Name}               & \textbf{Young's Modulus} & \textbf{Length} & \textbf{Height } & \textbf{Thickness} \\ \hline\hline
		ABCD                 & Base Plate                  & \SI{18.68}{MPa}          &\SI{ 3}{mm}      & \SI{1  }{mm}  & \SI{2}{mm}    \\ \hline
		M$_1$M$_2$M$_3$M$_4$ & Mechanism Domain            & \SI{18.68}{ MPa}        & \SI{2}{mm}       & \SI{1  }{mm}  & \SI{2}{mm}    \\ \hline
		P$_1$P$_2$P$_3$P$_4$ & Flexible Pillar             & \SI{18.68} {MPa }       & \SI{0.2}{mm}     & \SI{0.8}{mm} & \SI{0.2}{mm}  \\ \hline
		P$_5$P$_6$P$_7$P$_8$ & Flexible Pillar             & \SI{18.68 }{MPa }        & \SI{0.2}{mm}    & \SI{0.8}{mm}   & \SI{0.2}{mm} \\ \hline
		B$_1$B$_2$P$_8$P$_3$ & Biological Tissue Construct & \SI{0.20}{ MPa }         & \SI{1}{mm}       & \SI{0.1}{mm}  & \SI{0.2}{mm}  \\ \hline
	\end{tabular}
	\caption{Nomenclature, Material and dimensional specifications for the flexible poles design domain (Fig.~\ref{fig:ActualDesignDomain})} \label{Tab:ActualBioTissue}
\end{table*}
\begin{figure*}[h!]
	\begin{subfigure}[t]{0.3\textwidth}
		\centering
		\includegraphics[width=\linewidth]{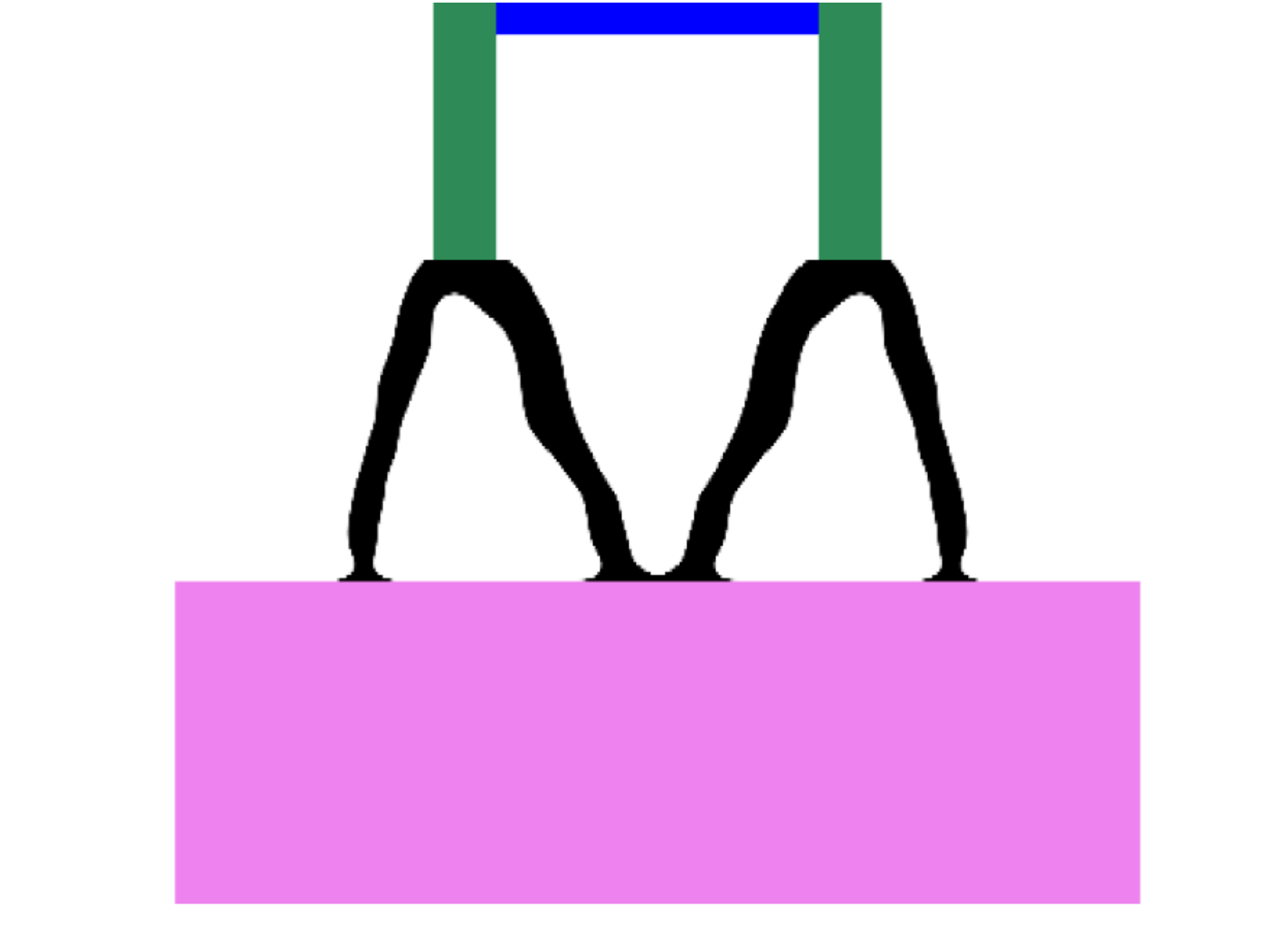}
		\caption{Intermediate design}
		\label{ActualSolutions5i}
	\end{subfigure}
	\begin{subfigure}[t]{0.3\textwidth}
		\centering
		\includegraphics[width=\linewidth]{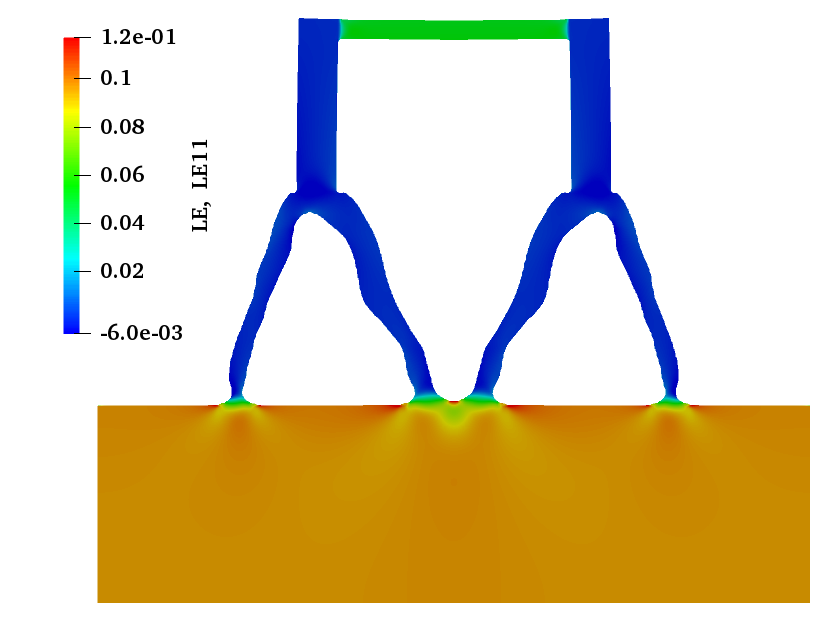}
		\caption{Strain distribution}
		\label{Xrobuststrain5}
	\end{subfigure}
	\begin{subfigure}[t]{0.3\textwidth}
		\centering
		\includegraphics[width=\linewidth]{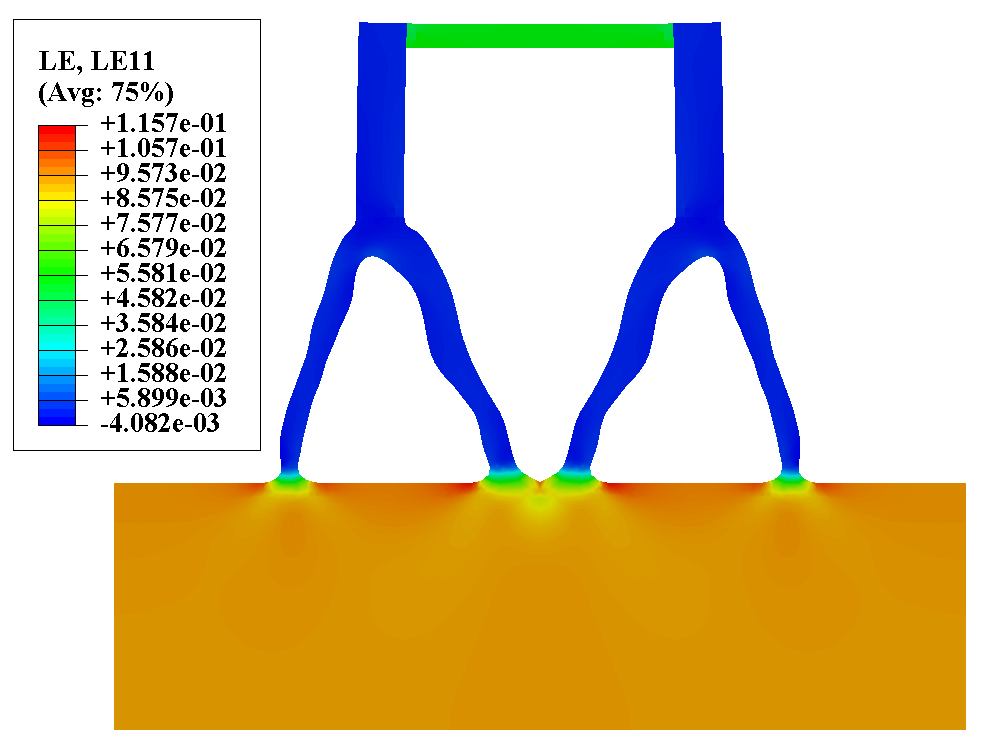}
		\caption{ABAQUS strain distribution}
		\label{ABAQUSstrain5}
	\end{subfigure}
	\begin{subfigure}[t]{0.475\textwidth}
		\centering
		\includegraphics[width=0.96\linewidth]{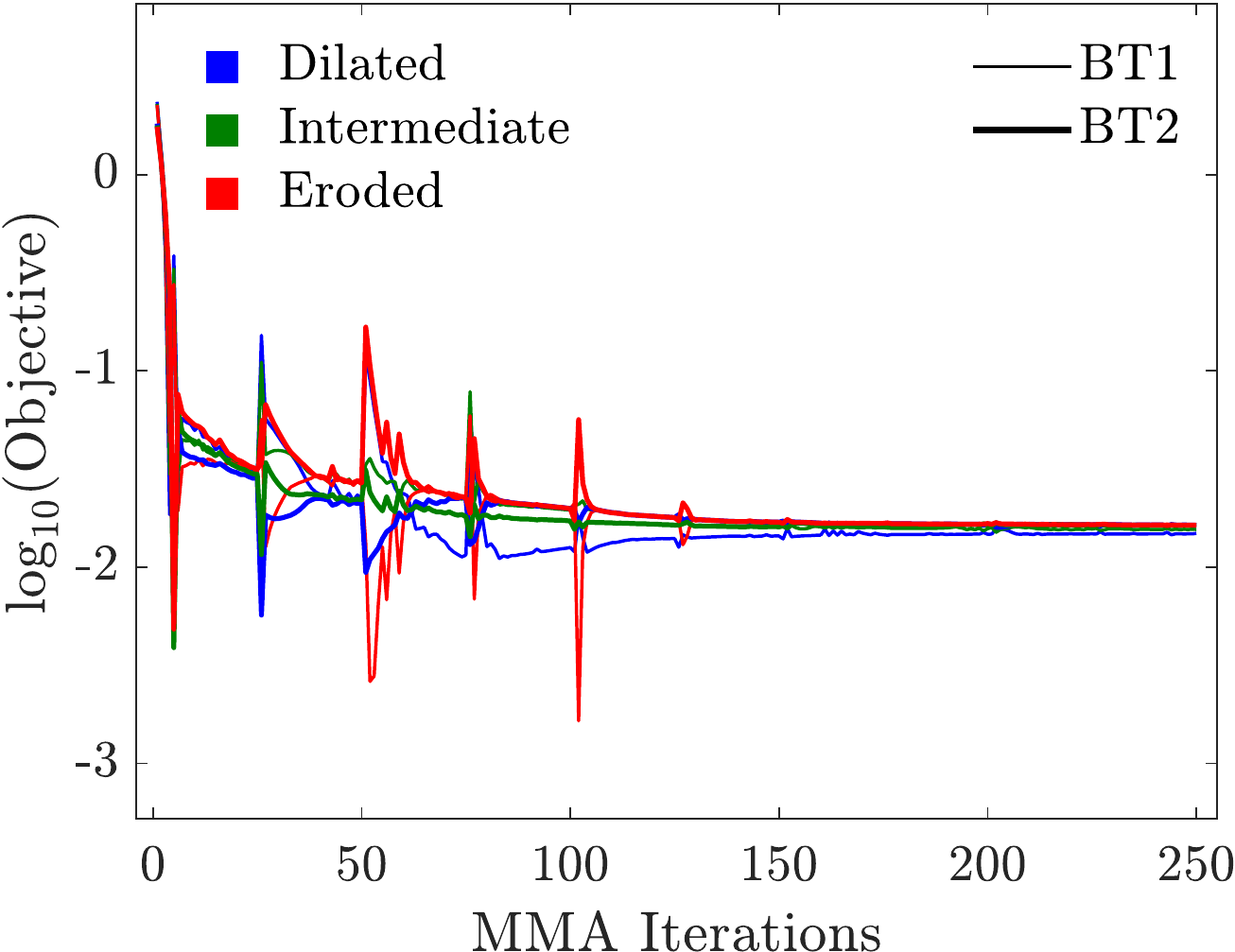}
		\caption{Objectives convergence plot}
		\label{ActualConverObjective5}
	\end{subfigure}
	\begin{subfigure}[t]{0.475\textwidth}
		\centering
		\includegraphics[width=\linewidth]{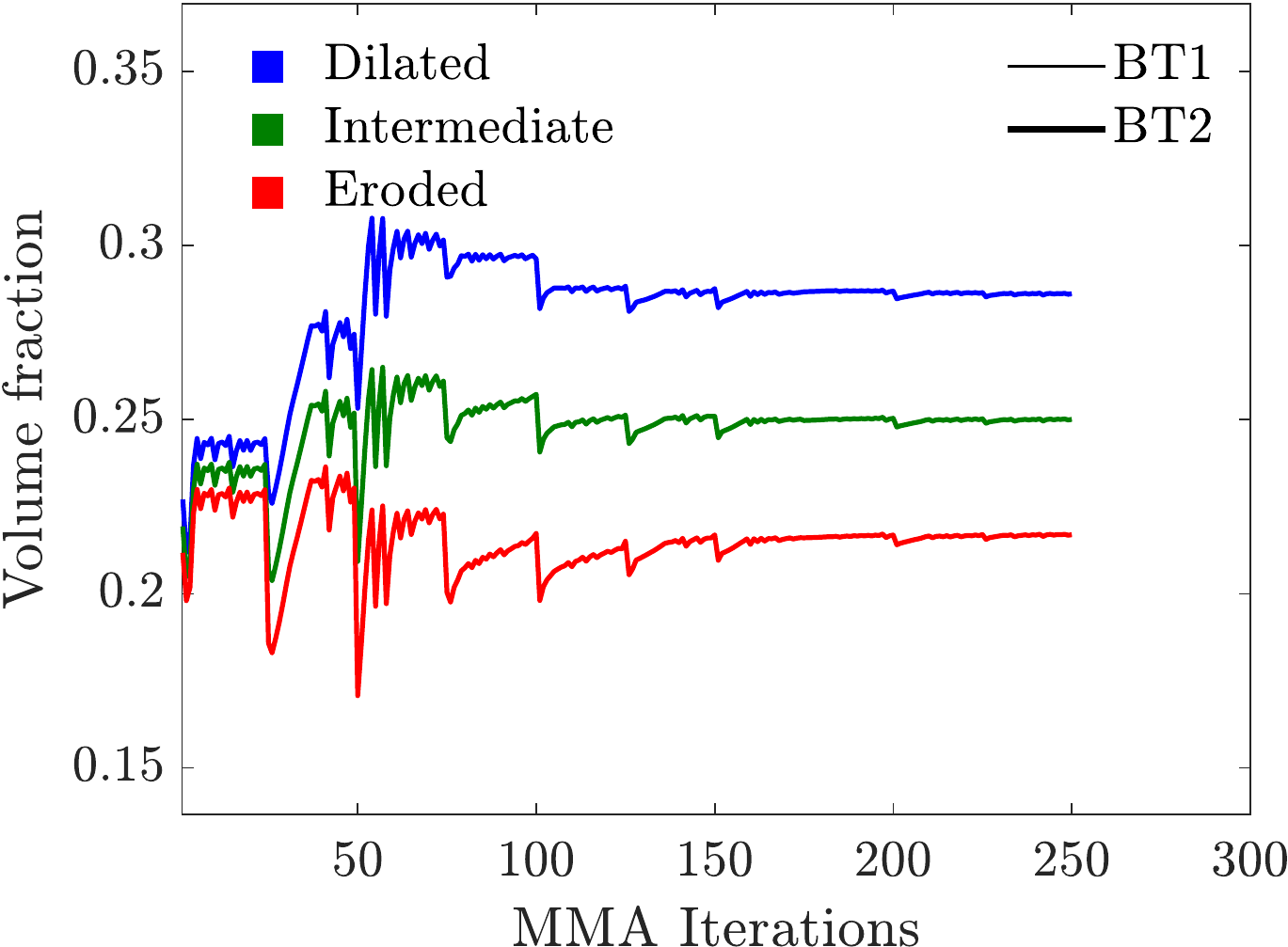}
		\caption{Volume constraints convergence plot}
		\label{ActualConverVolume5}
	\end{subfigure}
	\caption{Full solutions to CBM I (5\% desired straining). (\subref{ActualSolutions5i}) Optimized intermediate design, $M_\text{nd}=1.06\%$, (\subref{Xrobuststrain5}) Strain distribution obtained via the suggested approach, (\subref{ABAQUSstrain5}) Strain distribution obtained via ABAQUS analysis, (\subref{ActualConverObjective5}) Convergence plot for the objectives and (\subref{ActualConverVolume5}) Convergence plot for the volume fraction. Key: BT1: Actual biological tissue,\, BT2: Second biological tissue. }\label{fig:ActualSolutions5}
\end{figure*}
\begin{figure*}[h!]
	\begin{subfigure}[t]{0.3\textwidth}
		\centering
		\includegraphics[width=\linewidth]{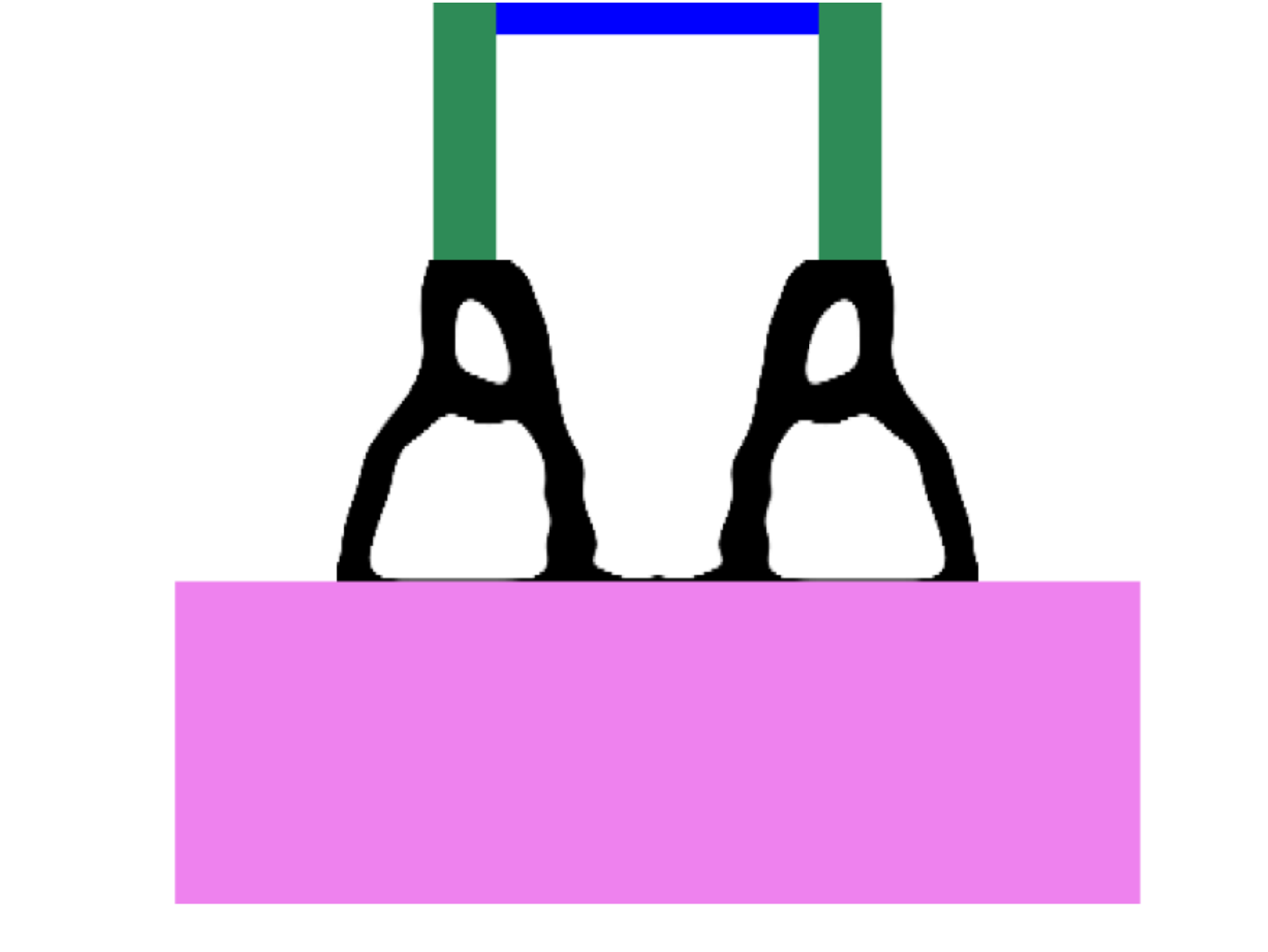}
		\caption{Dilated design}
		\label{ActualSolutions10d}
	\end{subfigure}
	\begin{subfigure}[t]{0.3\textwidth}
		\centering
		\includegraphics[width=\linewidth]{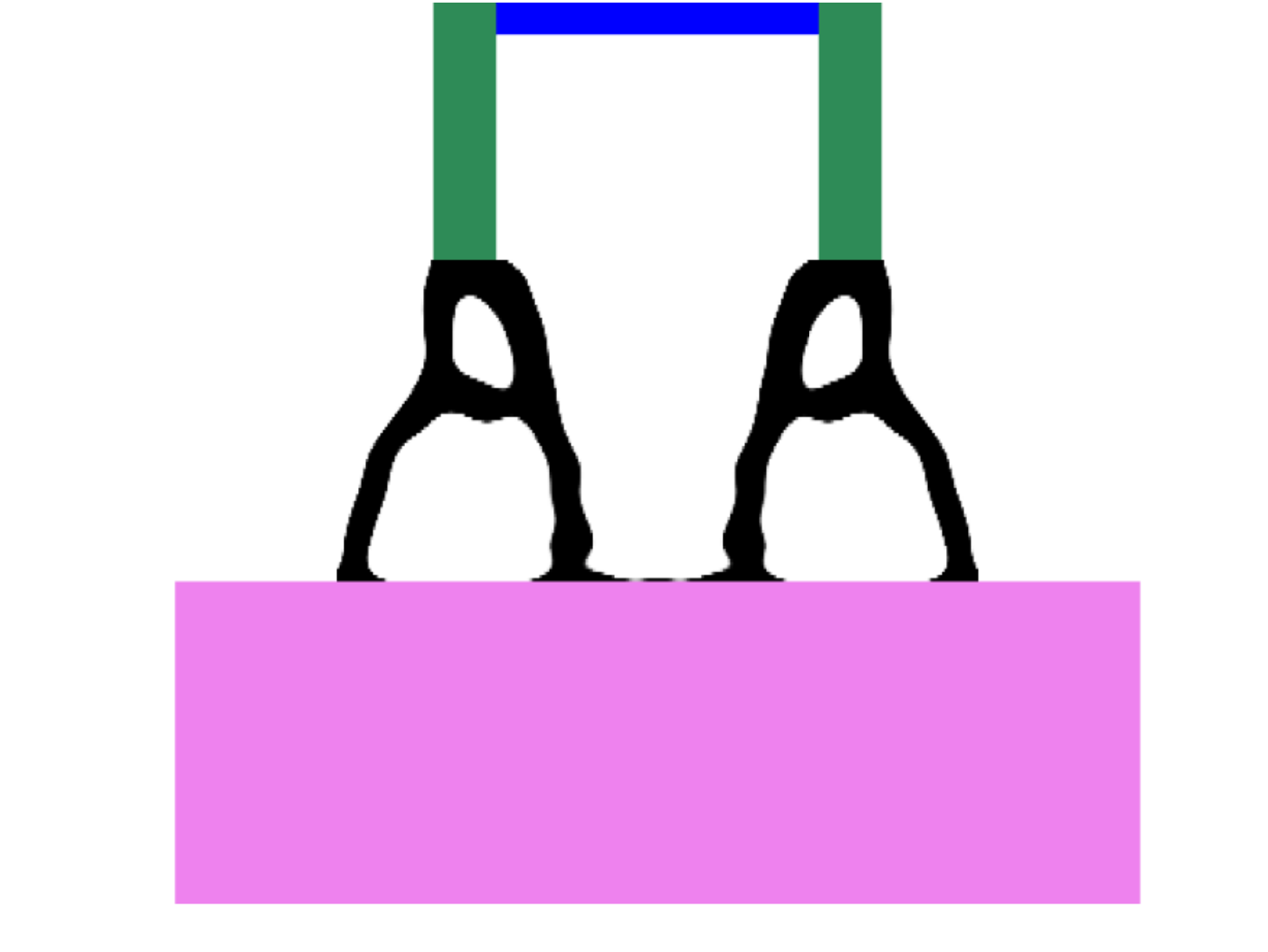}
		\caption{Intermediate design}
		\label{ActualSolutions10i}
	\end{subfigure}
	\begin{subfigure}[t]{0.3\textwidth}
		\centering
		\includegraphics[width=\linewidth]{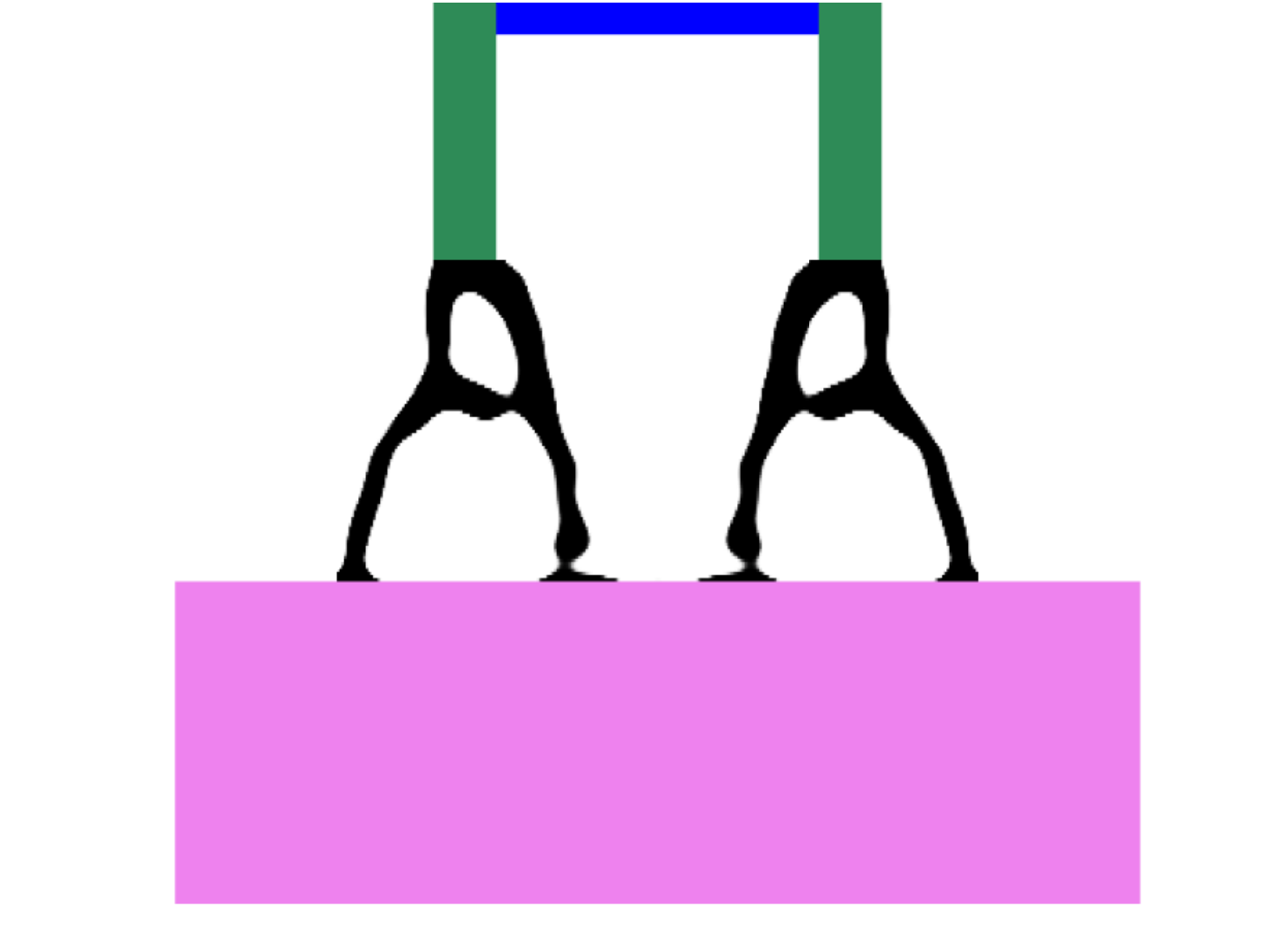}
		\caption{Eroded design}
		\label{ActualSolutions10e}
	\end{subfigure}
	\begin{subfigure}[t]{0.475\textwidth}
		\centering
		\includegraphics[width=\linewidth]{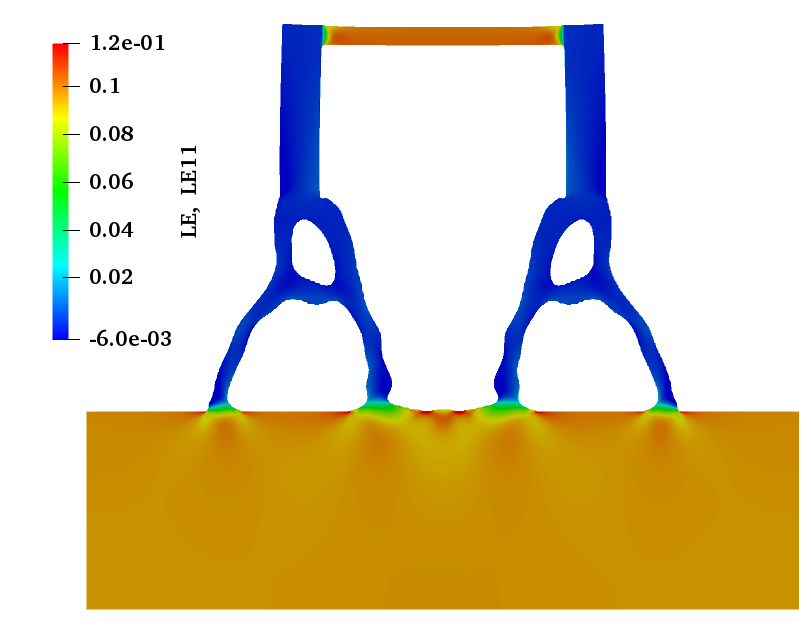}
		\caption{Strain distribution}
		\label{Xrobuststrain10}
	\end{subfigure}
	\begin{subfigure}[t]{0.475\textwidth}
		\centering
		\includegraphics[width=\linewidth]{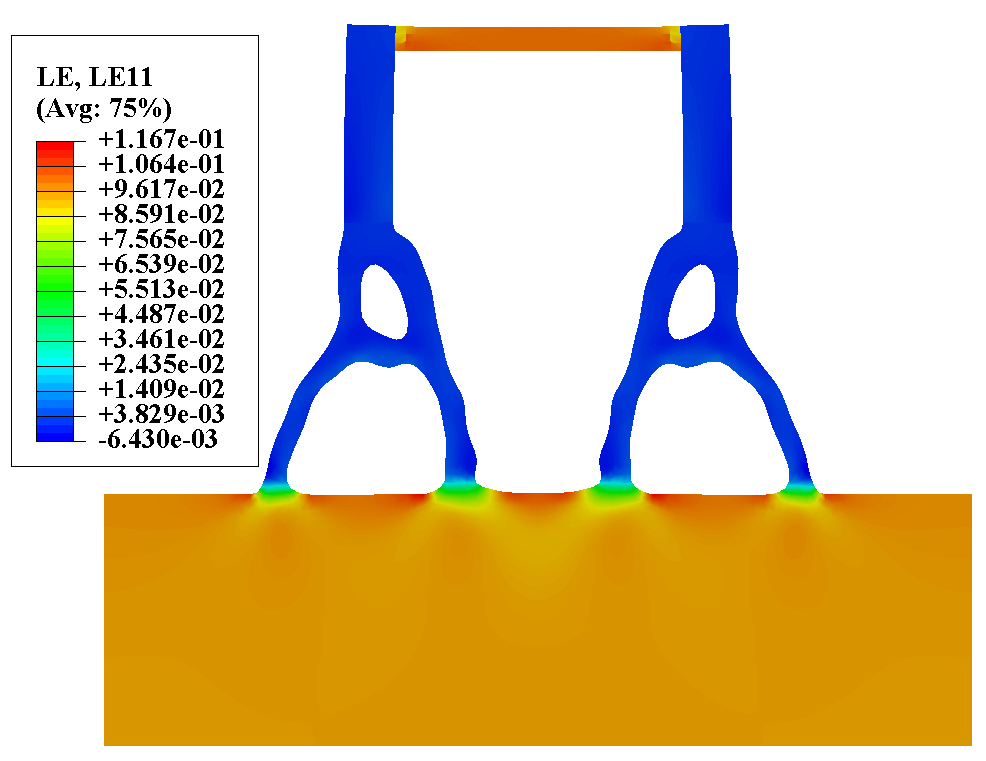}
		\caption{ABAQUS strain distribution}
		\label{ABAQUSstrain10}
	\end{subfigure}
	\caption{Full solutions to CBM II (10\% desired straining). (\subref{ActualSolutions10d}) Optimized dilated design, $M_\text{nd}=1.62\%$ (\subref{ActualSolutions10i}) Optimized intermediate design, $M_\text{nd}=1.35\%$, (\subref{ActualSolutions10e}) Optimized eroded design, $M_\text{nd}=1.56\%$, (\subref{Xrobuststrain10}) Strain distribution obtained via the presented approach and (\subref{ABAQUSstrain10}) Strain distribution obtained via ABAQUS analysis.}\label{fig:ActualSolutions10}
\end{figure*}

\subsection{SET-2: Numerical Example}\label{Sec:SET2}

The robustness of the synthesis approach using the  numerical examples is illustrated in Sec.~\ref{Sec:SET-1: Numerical Example}. This section presents a set of compliant actuators designed  in accordance with the \textit{flexible poles} method \citep{vandenburgh2008drug}. Herein, it is envisioned that the presented approach shall not only support one pair of flexible poles (one bioreactor) but can also provide support to many other similar bioreactors requiring different induced strains in biological tissues having either same or different geometries (see Fig.~\ref{fig:3DPic}). In Fig.~\ref{fig:3DPic}, different regions are indicated. For each pair of flexible poles, we seek the optimized compliant mechanisms which can provide  a specific desired strain $^n\epsilon_{xx}^*$ in the $n^\text{th}$ biological tissue when the base plate is strained uniformly by~$\Delta$.

\begin{figure*}[h!]
	\begin{subfigure}[t]{0.3\textwidth}
		\centering
		\includegraphics[width=\linewidth]{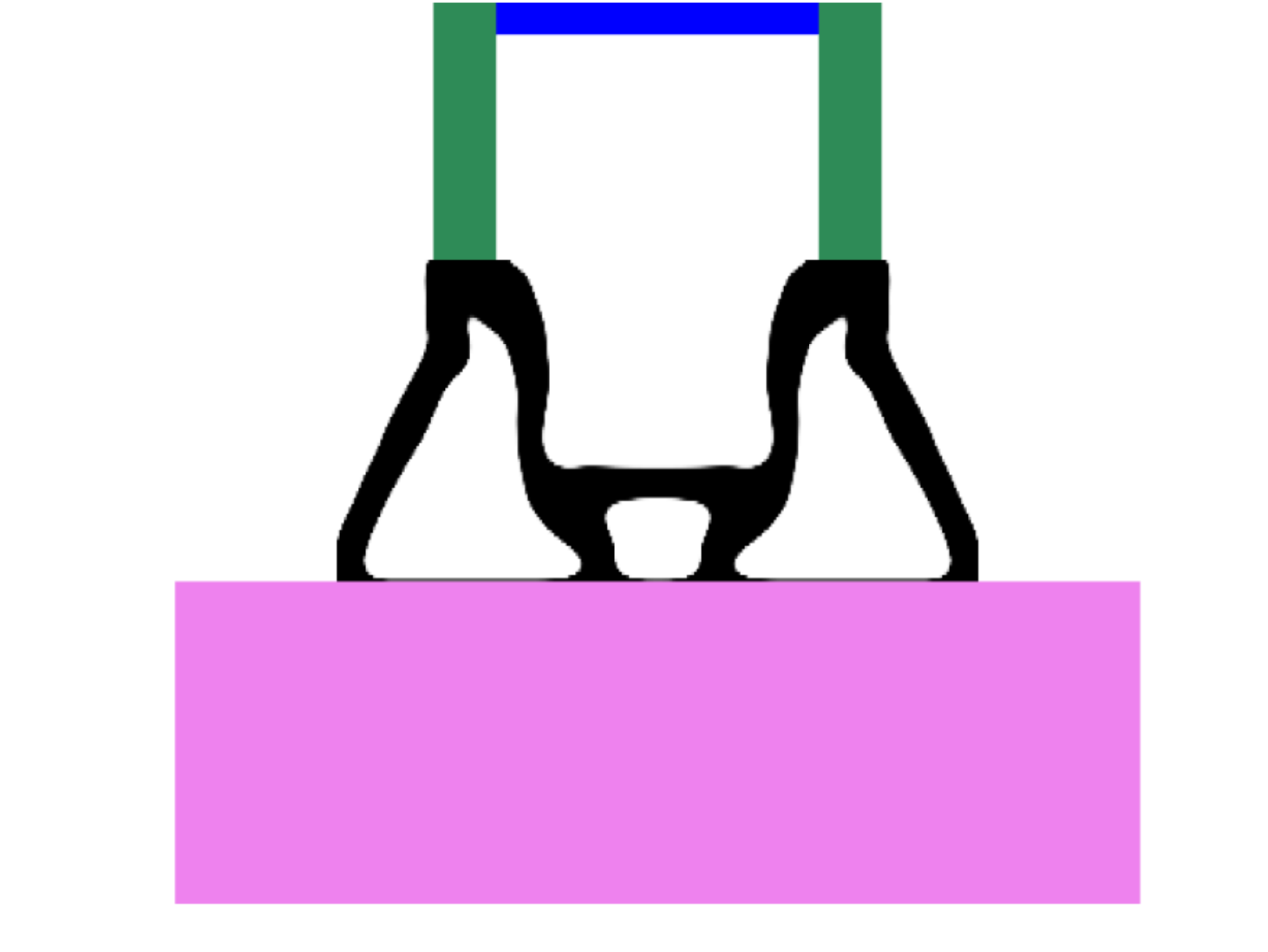}
		\caption{Dilated design}
		\label{ActualSolutions15d}
	\end{subfigure}
	\begin{subfigure}[t]{0.3\textwidth}
		\centering
		\includegraphics[width=\linewidth]{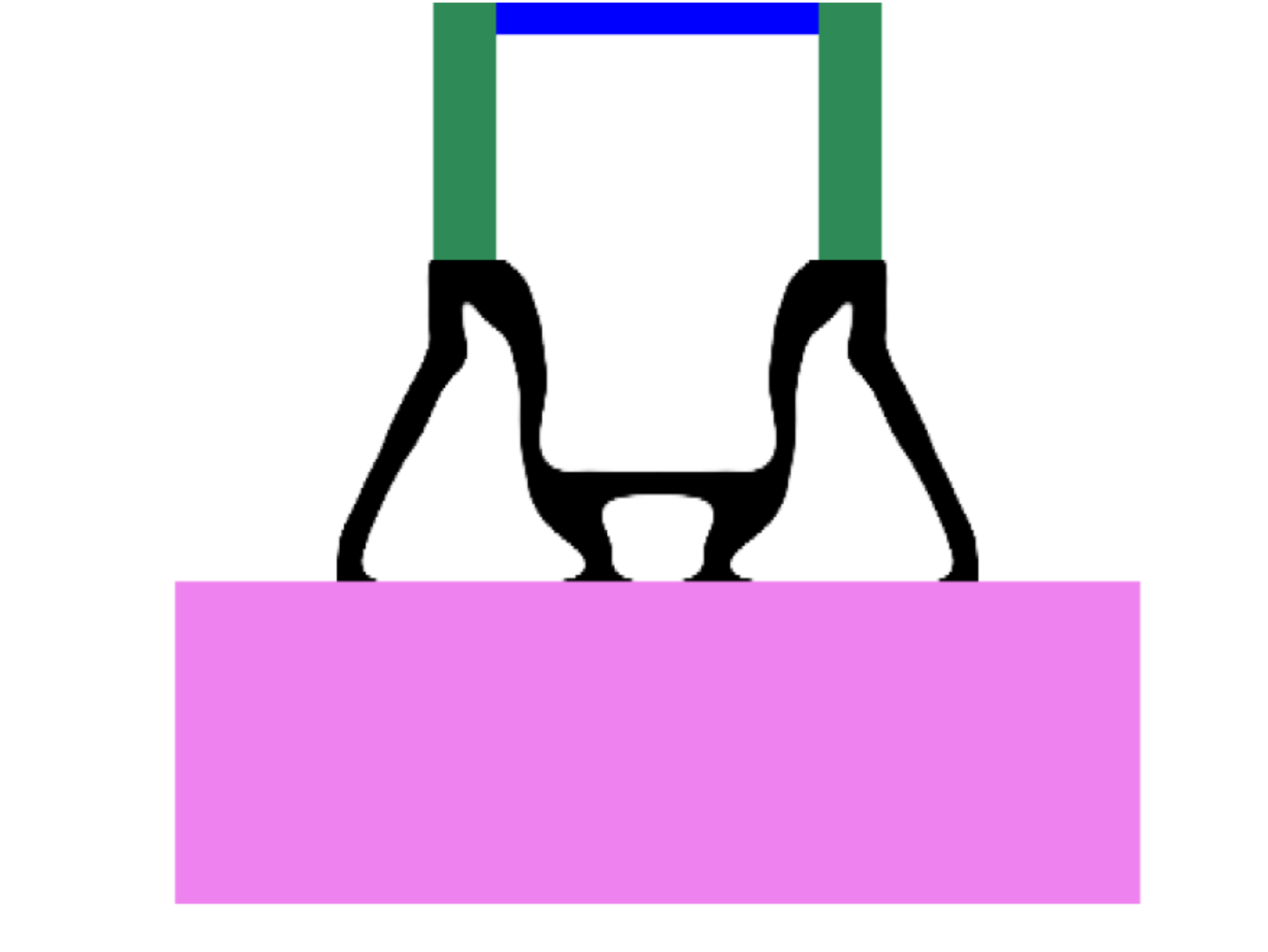}
		\caption{Intermediate design}
		\label{ActualSolutions15i}
	\end{subfigure}
	\begin{subfigure}[t]{0.3\textwidth}
		\centering
		\includegraphics[width=\linewidth]{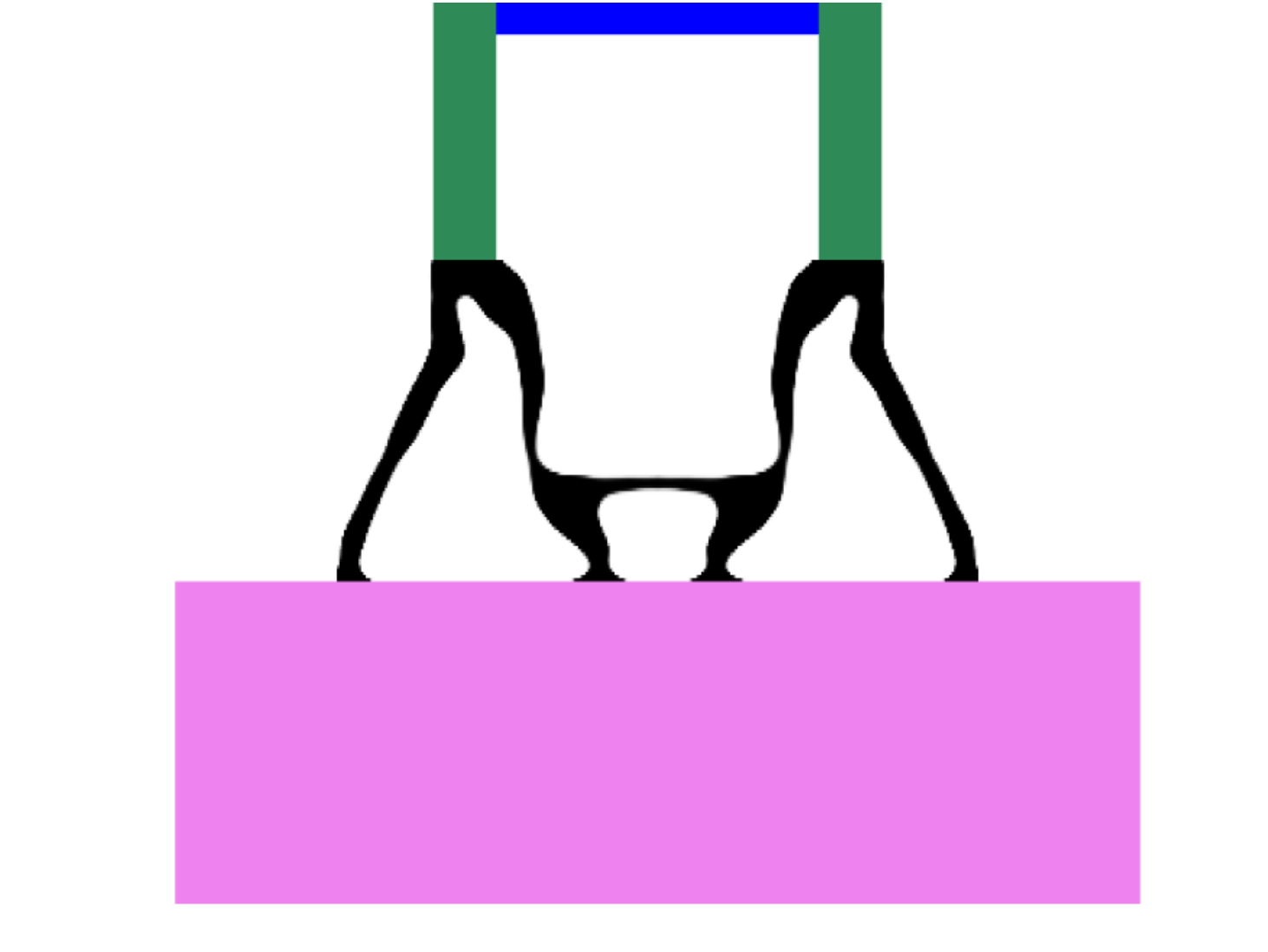}
		\caption{Eroded design}
		\label{ActualSolutions15e}
	\end{subfigure}
	\begin{subfigure}[t]{0.475\textwidth}
		\centering
		\includegraphics[width=\linewidth]{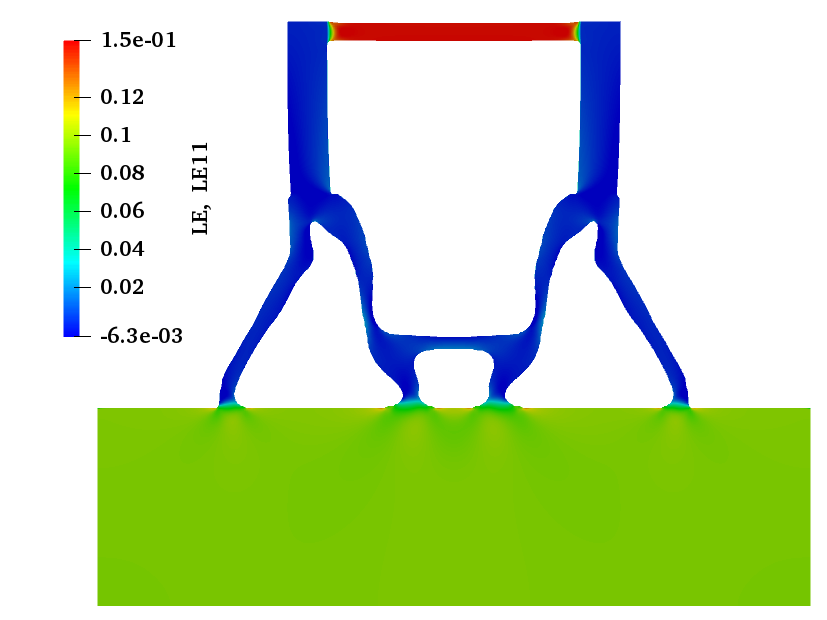}
		\caption{Strain distribution}
		\label{Xrobuststrain15}
	\end{subfigure}
	\begin{subfigure}[t]{0.475\textwidth}
		\centering
		\includegraphics[width=\linewidth]{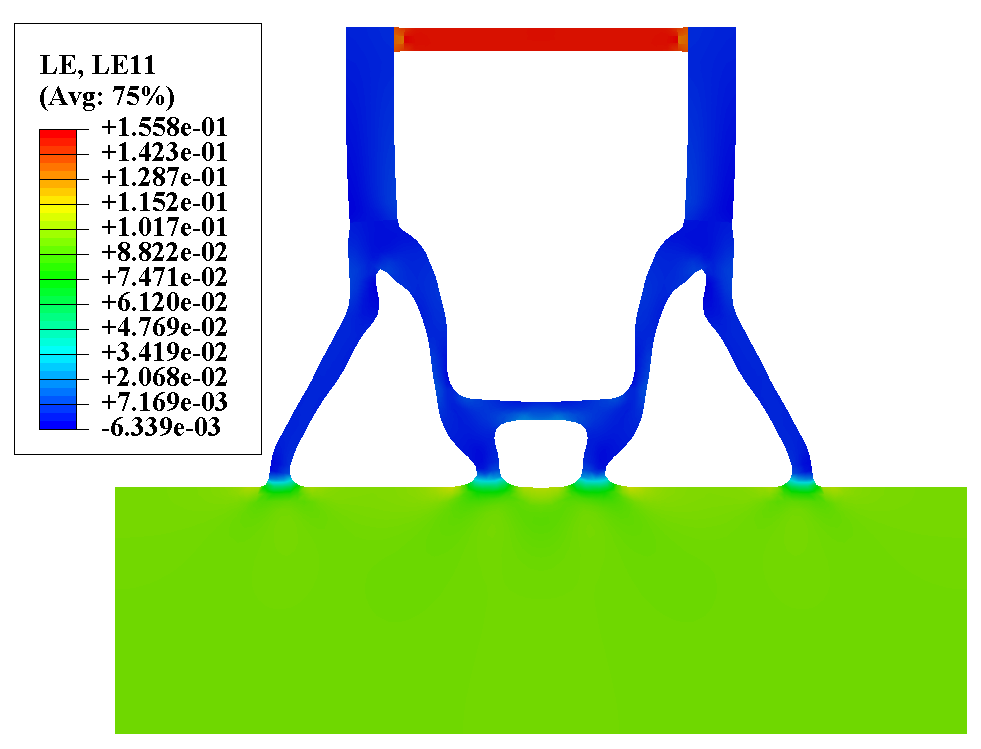}
		\caption{ABAQUS strain distribution}
		\label{ABAQUSstrain15}
	\end{subfigure}
	\caption{Full solutions to CBM III (15\% desired straining). (\subref{ActualSolutions15d}) Optimized dilated design, $M_\text{nd}=1.74\%$, (\subref{ActualSolutions15i}) Optimized intermediate design, $M_\text{nd}=1.27\%$, (\subref{ActualSolutions15e}) Optimized eroded design, $M_\text{nd}=1.70\%$, (\subref{Xrobuststrain15}) Strain distribution obtained via the suggested approach and (\subref{ABAQUSstrain15}) Strain distribution obtained via ABAQUS analysis.}\label{fig:ActualSolutions15}
\end{figure*}   
\begin{figure*}[h!]
	\begin{subfigure}[t]{0.3\textwidth}
		\centering
		\includegraphics[width=\linewidth]{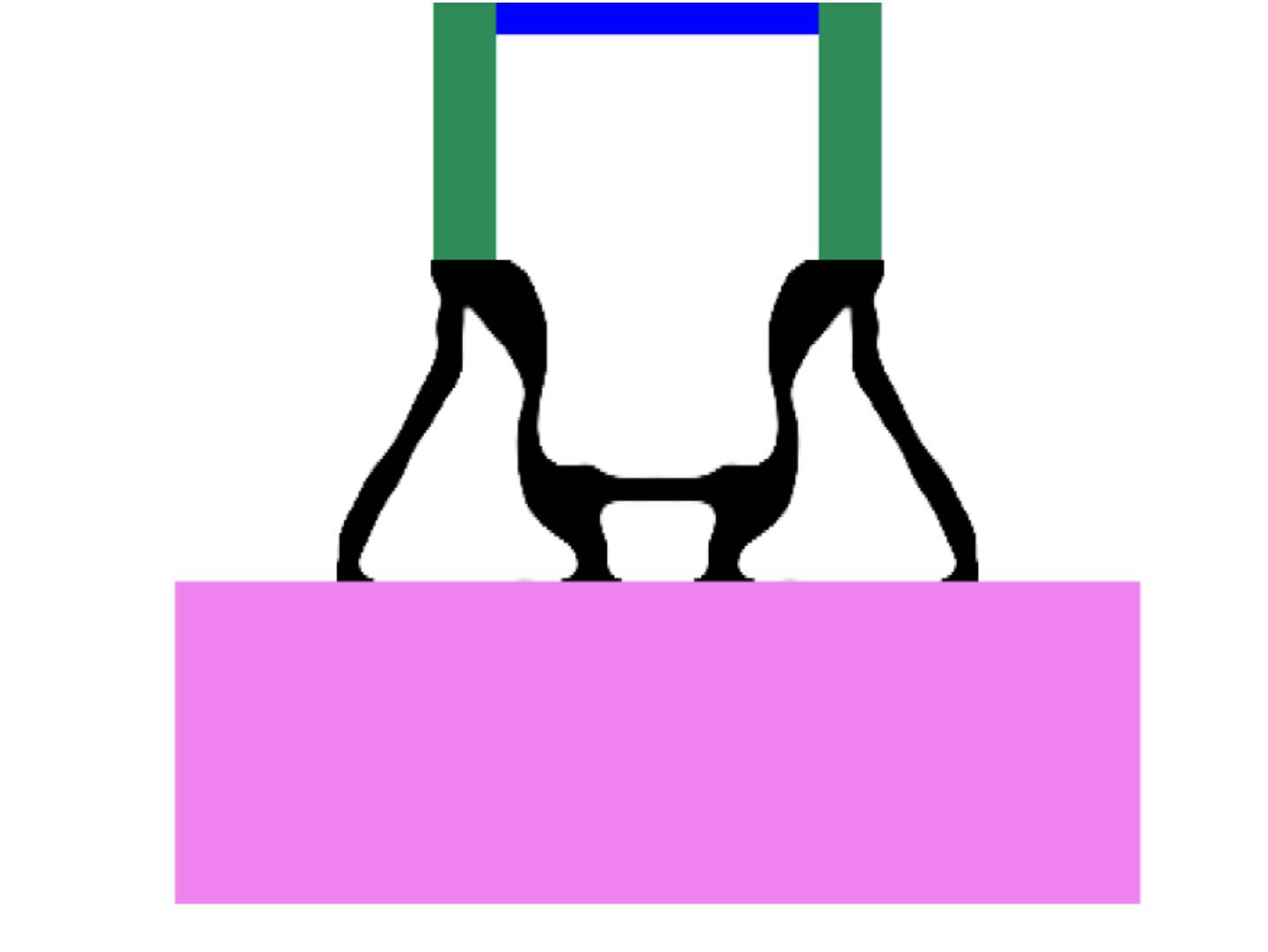}
		\caption{Intermediate design}
		\label{ActualSolutions20i}
	\end{subfigure}
	\begin{subfigure}[t]{0.30\textwidth}
		\centering
		\includegraphics[width=\linewidth]{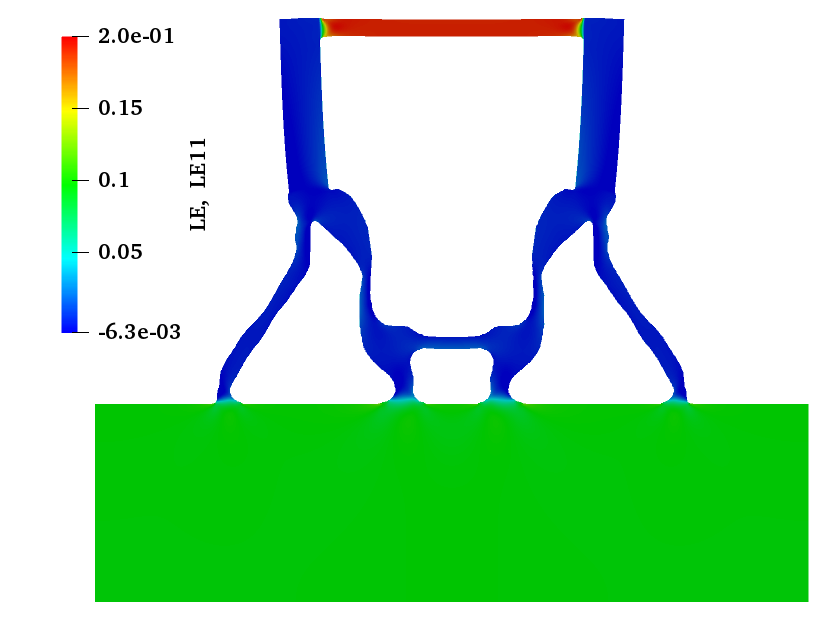}
		\caption{Strain distribution}
		\label{Xrobuststrain20}
	\end{subfigure}
	\begin{subfigure}[t]{0.3\textwidth}
		\centering
		\includegraphics[width=\linewidth]{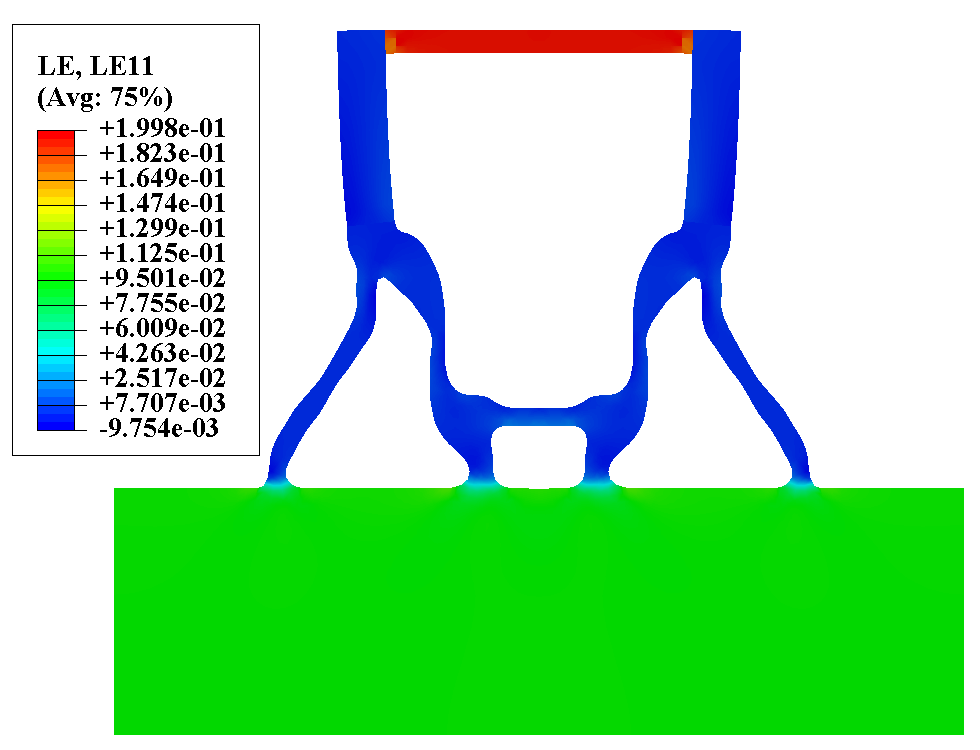}
		\caption{ABAQUS strain distribution}
		\label{ABAQUSstrain20}
	\end{subfigure}
	\begin{subfigure}[t]{0.475\textwidth}
		\centering
		\includegraphics[width=0.96\linewidth]{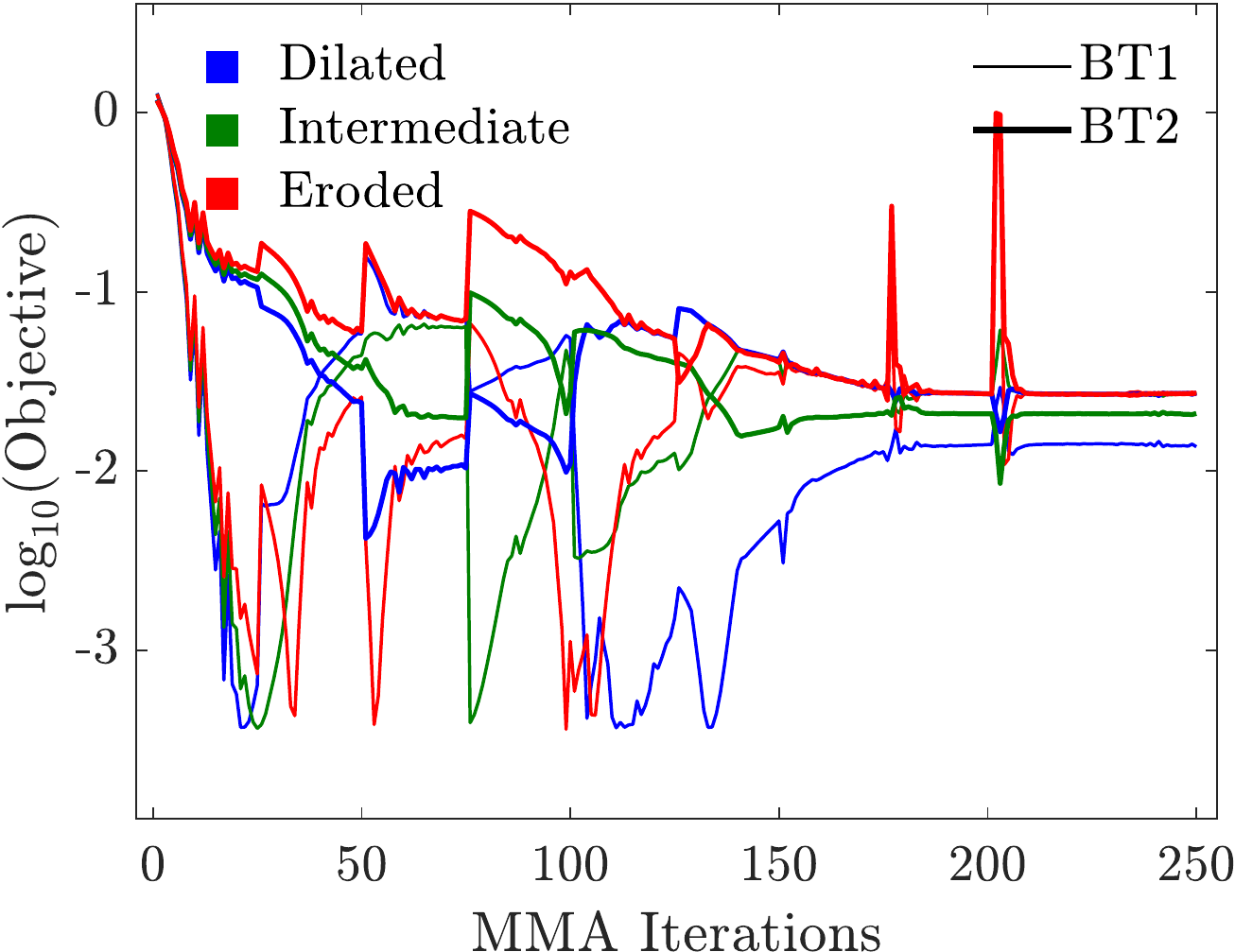}
		\caption{}
		\label{ActualConverObjective20}
	\end{subfigure}
	\begin{subfigure}[t]{0.475\textwidth}
		\centering
		\includegraphics[width=\linewidth]{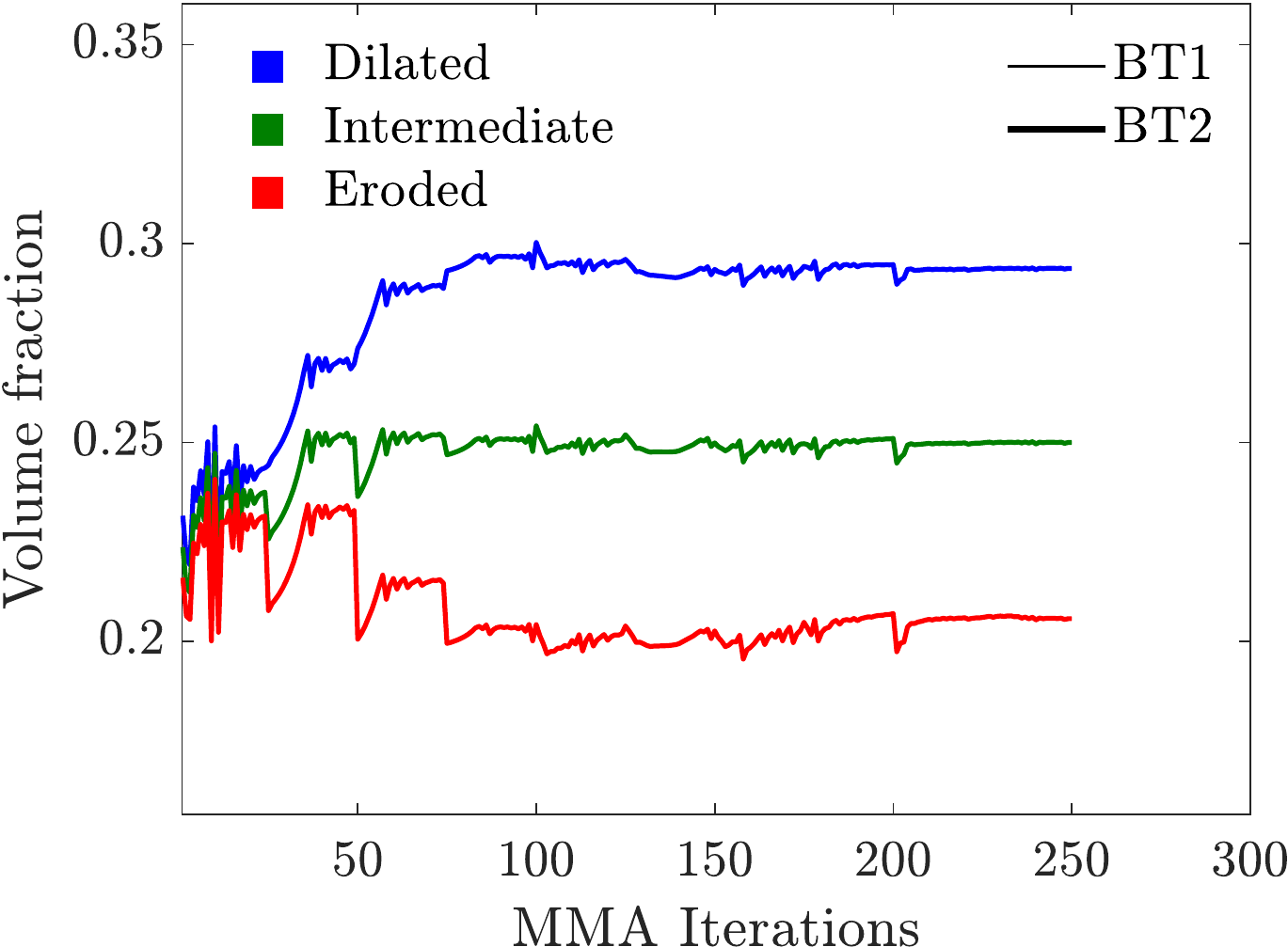}
		\caption{}
		\label{ActualConverVolume20}
	\end{subfigure}
	\caption{Full solutions to CBM IV (20\% desired straining). (\subref{ActualSolutions20i}) Optimized intermediate design, $M_\text{nd}=1.30\%$, (\subref{Xrobuststrain20}) Strain distribution obtained via the presented approach, (\subref{ABAQUSstrain20}) Strain distribution obtained via ABAQUS analysis, (\subref{ActualConverObjective20}) Convergence plot for the objectives and (\subref{ActualConverVolume20}) Convergence plot for the volume fraction. {Key:} BT1: Actual biological tissue,\, BT2: Second biological tissue.}\label{fig:ActualSolutions20}
\end{figure*}
Figure~\ref{fig:ActualDesignDomain} indicates the design domain specifications for designing compliant micro-actuator mechanisms with two flexible poles, a sample of biological tissue construct and a base plate. Table~\ref{Tab:ActualBioTissue} depicts the dimensions, material parameters (Young's moduli) and thicknesses  for these domains \citep{Pless3D}. The mechanisms are designed for achieving 5\%, 10\%, 15\% and 20\% straining in their respective biological tissue and named herein as \textit{compliant bio-mechanism} CBM I, CBM II, CBM III and CBM IV, respectively. The  color scheme of Fig.~\ref{fig:ActualDesignDomain} is used further to show the results wherein black color is used for the optimized mechanisms. 

  \begin{figure}[h!]
  	\begin{subfigure}[t]{0.45\textwidth}
  		\centering
  		\includegraphics[width=\linewidth]{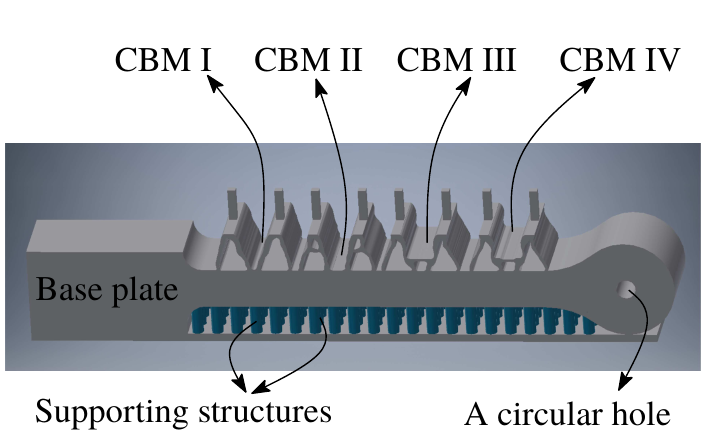}
  		\caption{CAD model}
  		\label{Experiment:CADModel}
  	\end{subfigure}
  	\begin{subfigure}[t]{0.45\textwidth}
  		\centering
  		\includegraphics[width=\linewidth]{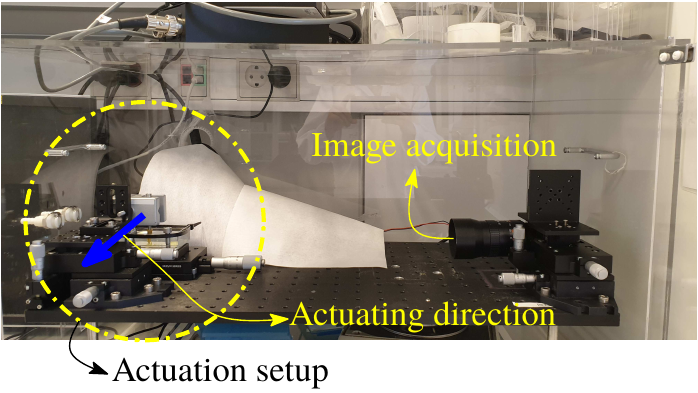}
  		\caption{Experimental setup}
  		\label{Experiment:Experimental setup}
  	\end{subfigure}
  	\begin{subfigure}[t]{0.45\textwidth}
  		\centering
  		\includegraphics[width=\linewidth]{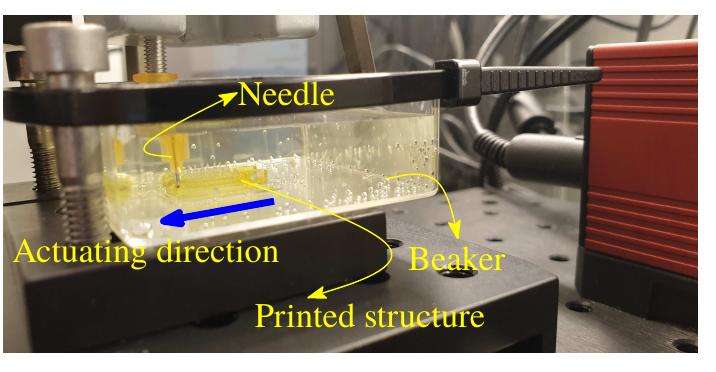}
  		\caption{Sample setup}
  		\label{Experiment:Sample setup}
  	\end{subfigure}
  	\caption{ (\subref{Experiment:CADModel}) 3D CAD model, (\subref{Experiment:Experimental setup}) Overview of the experimental setup and camera position and (\subref{Experiment:Sample setup}) Close-up of sample fixation}\label{fig:Experimentalsetup}
  \end{figure}

 Designing these robust compliant mechanisms poses unique challenges because: (i) the biological tissue is comparatively very soft, (ii) the tissue construct is connected\footnote{Not directly actuated by the mechanism} to the micro-mechanism via flexible poles and (iii) it is essential to consider geometric  nonlinearity as already mentioned before. To account for the challenges and also to permit different tissue construct geometries, we use the extended robust formulation with $N_\text{BT}  =2$ (see Sec.~\ref{Sec:OptimizationProblemformulation}). The first tissue construct corresponds to the actual (minimum) given size, whereas the width of the second tissue construct BT2 is assumed to be two times that of the first one ensuring tissues with different (higher) stiffness.
  
The symmetric half of the design domain ABEF (Fig.~\ref{fig:ActualDesignDomain}) is discretized using $N_\text{ex}\times N_\text{ey}~=200\times280$ quad-FEs, and thereafter, FEs representing different regions are detected.  FEs describing  the base plate, poles and biological tissue are attributed by $\rho=1$ throughout the optimization process. Likewise,  $\rho=0$ is designated to all FEs associated to the void regions. Roller boundary conditions permitting movement of the bottom and left sides of the plate along $x-$ and $y-$axes, respectively, are considered, whereas the right side of the plate is actuated by an amount $\Delta = 0.1L_x$, i.e., 10\% of uniform straining in the base plate (Fig.~\ref{fig:ActualDesignDomain}). Filter radius is set to $10\times\max(\frac{L_x}{2N_\text{ex}}, \frac{Ly}{N_\text{ey}}) $. The Heaviside projection filter parameter $\beta$ is altered from 1 to 128. It is double at each $25^\text{th}$ MMA iteration till it reaches to 128 and thereafter, it remains 128 for the remaining optimization iterations. The maximum number of optimization iterations is set to 250. The optimized results are shown with the actual tissue construct.

Figure~\ref{ActualSolutions5i}, Fig.~\ref{ActualSolutions10i}, Fig.~\ref{ActualSolutions15i}, and Fig.~\ref{ActualSolutions20i} show the optimized intermediate designs for CBM~I, CBM~II, CBM~III and CBM~IV respectively. Figure~\ref{ActualSolutions10d} and Fig.~\ref{ActualSolutions15d},  and  Fig.~\ref{ActualSolutions10e} and Fig.~\ref{ActualSolutions15e}  depict optimized dilated and eroded designs for the CBM~II and CBM~III. One notices the optimized topologies CBM~III and CBM~IV are identical, however the latter one has  comparatively some thin slender sections facilitating more deformation and thus, help providing close to 20\% strain  in the tissue construct (Fig.~\ref{Xrobuststrain20} and Fig.~\ref{ABAQUSstrain20}). The objectives convergence plots for CBM~I and CBM~IV are indicated by Fig.~\ref{ActualConverObjective5} and Fig.~\ref{ActualConverObjective20} respectively and their corresponding volume fraction convergence plots are depicted via Fig.~\ref{ActualConverVolume5} and Fig.~\ref{ActualConverVolume20}.

As $\beta$ changes, the approximating function in Eq.~\ref{eq:Heaviside} alters and thus, as per $\beta$ continuation, jumps in the convergence curves of the objectives and volume constraints can be noticed. In addition, with updates in the volume of the dilated design, the convergence curves may also get altered. Convergence plots are smooth after $200$ MMA optimization iterations and volume constraint is satisfied and remains active in the end of the optimization (Figs.~\ref{ActualConverObjective5}, \ref{ActualConverVolume5},\,\ref{ActualConverObjective20},\, and \ref{ActualConverObjective20}).

 One can notice uniform distributions of actual strains, close to their desired ones, within the respective tissues of CBM~I, CBM~II, CBM~III and CBM~IV (Fig.~\ref{Xrobuststrain5}, Fig.~\ref{Xrobuststrain10}, Fig.~\ref{Xrobuststrain15} and Fig.~\ref{Xrobuststrain20}). The actual strain distribution is demonstrated with respect to the optimized intermediate designs. The strain errors Err$_x$ (Eq.~\ref{eq:errormeasure}) $12.4\%$, $13.0\%$, $14.4\%$ and $14.8\%$ are noticed for CBM~I, CBM~II, CBM~III and CBM~IV, respectively.

To demonstrate the accuracy of the obtained results, extracted and smoothed intermediate designs are also analyzed in ABAQUS with the same boundary conditions, actuating forces and material properties as those used by the formulation. Figure~\ref{Xrobuststrain5}, Fig.~\ref{Xrobuststrain10}, Fig.~\ref{Xrobuststrain15}, and Fig.~\ref{Xrobuststrain20} illustrate the actual strain distributions and their respective results obtained by ABAQUS analyses are depicted by Fig.~\ref{ABAQUSstrain5}, Fig.~\ref{ABAQUSstrain10}, Fig.~\ref{ABAQUSstrain15}, and Fig.~\ref{ABAQUSstrain20} respectively. One can see that the strain distributions obtained by the presented approach and ABAQUS analyses are in close agreement with each other.

\subsubsection{Prototypes of compliant micro-mechanisms and their performances}

A stereolithography-based 3D printing process \citep{zhang2017stereolithographic} is employed to fabricate the optimized mechanisms CBM~I, CBM~II, CBM~III and CBM~IV with their flexible poles and a base plate wherein a hydrogel material based on poly~(ethylene glycol) diacrylate is used for printing of the final prototypes. The adopted printing steps are as follows: 

\begin{figure}
	\centering
	\includegraphics[scale = 1]{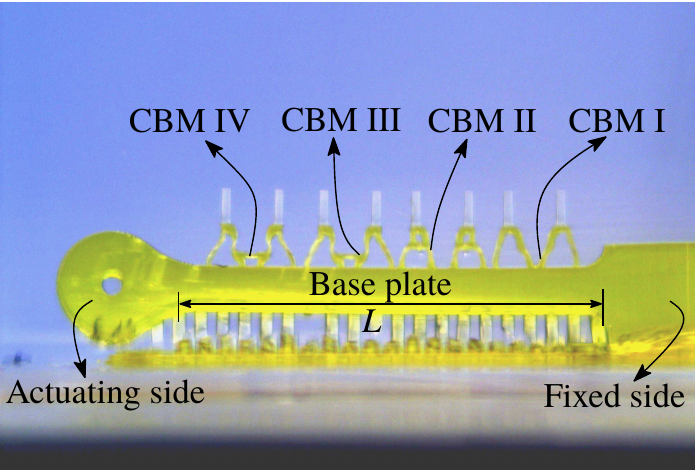}
	\caption{Photo of 3D printed mechanisms, immersed in water, with a base plate of length $L= 10.2$ mm and corresponding flexible poles}
	\label{fig:Experiment:Printed}
\end{figure}

\begin{enumerate}
	\item The optimized mechanisms are converted into  3D CAD (Computer-aided design) models with their flexible poles. These models are assembled on a base plate. A circular hole with diameter $\SI{500}{\micro \meter}$ is extruded from the one end of the base-plate to facilitate actuation/stretching (Fig.~\ref{Experiment:CADModel}). 
	\item Cylindrical supporting structures, radii $\SI{150}{\micro \meter}$ and heights $\SI{750}{\micro \meter}$, are placed on a square grid of center-to-center spacing $\SI{500}{\micro \meter}$, i.e., with a  $\SI{200}{\micro \meter}$  gap to ease the printing process (Fig.~\ref{Experiment:CADModel}).
	\item The assembled 3D CAD design is sliced into $\SI{20}{\micro \meter}$ layers using the open-source software Slic3r for printing process.
	\item 	A photo-curable resin consisting of 50 \%v/v poly~(ethylene glycol) diacrylate, M$_n$ 700 g/mol,  12 mg/mL Quinoline Yellow, and 5 mg/mL lithium phenyl-2,\,4,\,6-trimethylbenzoylphosphinate in water is selectively exposed to 365 nm ultraviolet light for 3 s per layer.
	\item 	Printing is conducted on a 3-(trimethoxysilyl)propyl methacrylate-treated cover-glass to ensure adhesion of the first layers to the printing platform.
	\item After printing, structures are washed in water and swollen to equilibrium. Support structures are detached from the print with the help of a needle. 
\end{enumerate}

\begin{figure}
	\begin{subfigure}[t]{0.45\textwidth}
		\centering
		\includegraphics[width=\linewidth]{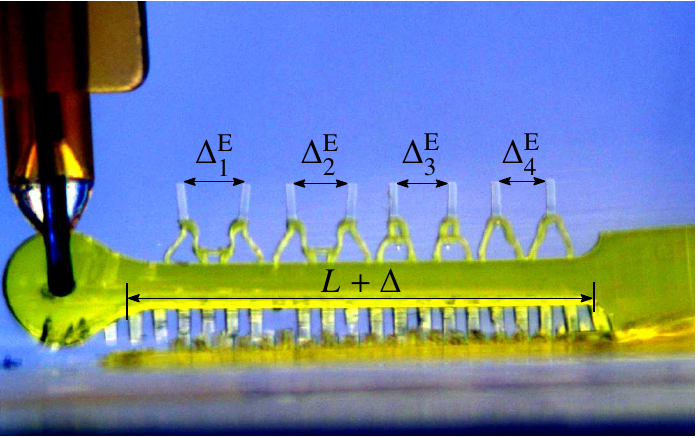}
		\caption{Deformed profiles: Experiment}
		\label{Experiment:DeformedMechanism}
	\end{subfigure}
	\begin{subfigure}[t]{0.45\textwidth}
		\centering
		\includegraphics[width=\linewidth]{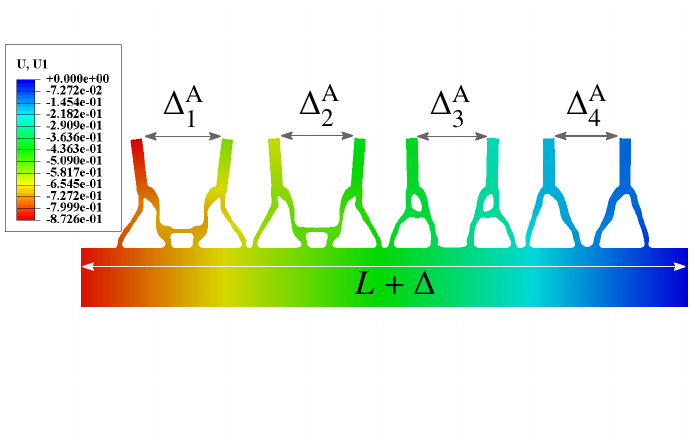}
		\caption{Deformed profiles: ABAQUS}
		\label{ABAQUS:DeformedMechanism}
	\end{subfigure}
	\caption{Deformed profiles of the mechanisms. (\subref{Experiment:DeformedMechanism}) Experimental deformed profiles with $\Delta_1^E = 0.32l_0,\,\Delta_1^E = 0.26l_0,\,\Delta_1^E = 0.14l_0,\,$ and $\Delta_1^E = 0.03l_0$ \subref{ABAQUS:DeformedMechanism} Deformed profiles obtained by ABAQUS analysis wherein $\Delta_1^A = 0.27l_0,\,\Delta_1^A = 0.20l_0,\,\Delta_1^A = 0.13l_0,\,$ and $\Delta_1^A = 0.063l_0$. Here, $l_0 =\SI{1}{mm}$ i.e. the initial gap between each pair of flexible poles and $\Delta=0.08L$.}\label{fig:ExpABAQUS}
\end{figure}
In the experimental setup (Fig.~\ref{fig:Experimentalsetup}), the cover-glass with attached printed structures is fixed on the bottom of a rectangular transparent beaker filled with water (Fig~\ref{Experiment:Sample setup}). The prototype is aligned by 3-axes stages (RB13M, Thorlabs, Inc.).  A $\SI{500}{\micro \meter}$ outer-diameter needle is fixed to the actuation stage and its tip is bent by $90^o$ and subsequently inserted into the hole on the actuating side. Actuation is carried out by moving the needle in steps of $\SI{100}{\micro \meter}$  in the actuation direction until failure $(\Delta)$. An image for every position is acquired by a camera (UI-3880LE-C-HQ, IDS Imaging) placed orthogonally to the actuated platforms (Fig~\ref{Experiment:Experimental setup}).

At the preliminary stage (presented herein), the experiment is performed without biological tissue constructs (Fig.~\ref{fig:Experimentalsetup}).  Figure~\ref{fig:Experiment:Printed} indicates prototypes of the mechanisms with their flexible poles and a base plate having length $L$. The base plate is stretched up to 8\% of its length i.e. $\Delta =0.08L$ (Fig.~\ref{Experiment:DeformedMechanism} and Fig.~\ref{ABAQUS:DeformedMechanism}). The deformed profiles of the mechanisms are imaged and displayed  in Fig.~\ref{Experiment:DeformedMechanism}. An animated image sequence of the gradual profile deformation as a function of stretching is included in the supplementary material (Movie~1). A corresponding ABAQUS model is developed and analyzed. Figure~\ref{ABAQUS:DeformedMechanism} illustrates the deformed profiles of the mechanisms with respective flexible poles obtained via the ABAQUS analysis. The printed mechanisms perform as they are expected, i.e, they could move apart their respective flexible poles and thus, can induce strains in the respective tissue constructs when base plate is actuated.  The tiny dimensions and compliant materials used do not allow for functional modeling using externally generated phantoms, such as suspended rubber bands. Functional testing with microtissues suspended between sets of poles is outside the scope of the current work. However, we have recently demonstrated such stable and reproducible muscle tissue generation in fully 3D printed flexible pole devices of similar dimensions assisted by an integral microreservoir for cell seeding \citep{christensen20193d}, which we will combine with the presented TO approach in future work.

The recorded axial stretching in the flexible poles pertaining to CBM~I, CBM~II, CBM~III and CBM~IV via the experiment are $\Delta_1^E = 0.03l_0$, $\Delta_1^E = 0.14l_0$, $\Delta_1^E = 0.26l_0$ and $\Delta_1^E = 0.32l_0$, respectively. Corresponding stretching obtained from the ABAQUS analysis are  $\Delta_1^A = 0.063l_0$, $\Delta_1^A = 0.13l_0$, $\Delta_1^A = 0.20l_0$ and $\Delta_1^A = 0.27l_0$, where $l_0=\SI{1}{mm}$. It can be noted that experimental result show: (i) higher deformation for CBM~III and CBM~IV, (ii) close agreement for CBM~II and (iii) lower deformation for CBM~I with respect to its ABAQUS analysis. There could be many reasons for such discrepancies, e.g., geometry variations, description of material properties including the Poisson's ratio, boundary conditions, out-of-plane bending of the 3D-printed flexible poles, and they need further and deeper investigations which are out of scope of the current manuscript and left for our future endeavors. 

	\section{Conclusions}\label{Sec:Conclusions}
This paper presents a method using topology optimization to design large deformation compliant mechanisms for inducing the desired strains in biological tissues. An objective  based on least square error is formulated using the given target strains and minimized. To cater
for large deformation and material properties of the
tissues, geometric and material nonlinearities through a suitable neo-Hookean material model are considered. The versatility of the presented approach is demonstrated by designing mechanisms which can induce strains in the biological tissues in both axial and bi-axial directions.

The mechanism design problem is conceptualized in the flexible poles environment, and various compliant mechanisms are successfully designed for inducing different strain levels in their respective biological tissues. The robust formulation is extended to accommodate different geometries of the tissue constructs. A base plate is used to actuate the mechanisms which render specific movements in their associated flexible poles and thus, help inducing the target strains in the tissues. Actual strain distributions in the tissues by the optimized mechanisms using the approach closely resemble those determined using their respective ABAQUS analyses.

The optimized mechanisms with their flexible poles and a base plate are 3D-printed using poly(ethylene glycol) diacrylate material and a simplified experiment is performed. With respect to its corresponding ABAQUS analysis, we observe good qualitative agreement with  some discrepancies in the stretches developed by their flexible poles. These discrepancies could have resulted from geometry variations, human errors, boundary conditions, material properties, and, a subject for our near future study. In addition, extension to a 3D setting with flexible poles environment is one of the prime directions for future work.  

	\section*{Acknowledgment}
	All authors acknowledge support from Independent Research Fund Denmark, grant 7017-00366B.  O. Sigmund acknowledges the support from the Villum Investigator project InnoTop provided by the  Villum Foundation. The authors also acknowledge Prof. Krister Svanberg for providing MATLAB codes of the MMA optimizer.
	\begin{appendices}
		\numberwithin{equation}{section}
		\section{Evaluating the derivative $\pd{f_k}{\mb{u}}$}\label{Append1}
In view of \eq{eq:objective_FE}, one finds the derivative $\pd{f_k}{\mb{u}_e}$ as\footnote{Subscript $k$ and term related to shear strain from the numerator of the objective are dropped for simplicity.}

\begin{equation}\label{eq:append_1}
\begin{aligned}
\pd{f}{\mb{u}_e} =\frac{2}{N_{be}}\displaystyle\sum_{e= 1}^{N_{be}}\left(\frac{w_1(\epsilon_\ms{xx}^e -\epsilon_\ms{xx}^*)\pd{\epsilon_\ms{xx}^e}{\mb{u}_e} +w_2(\epsilon_\ms{yy}^e -\epsilon_\ms{yy}^*)\pd{\epsilon_\ms{yy}^e}{\mb{u}_e} + +w_3(\epsilon_\ms{xy}^e -\epsilon_\ms{xy}^*)\pd{\epsilon_\ms{xy}^e}{\mb{u}_e}} {w_1(\epsilon_\ms{xx}^*)^2 + w_2(\epsilon_\ms{yy}^*)^2 + w_3(\epsilon_\ms{xy}^*)^2} \right)
\end{aligned}
\end{equation}

Therefore, one needs $\pd{\epsilon_\ms{xx}^e}{\mb{u}_e}$, $\pd{\epsilon_\ms{yy}^e}{\mb{u}_e}$ and $\pd{\epsilon_\ms{xy}^e}{\mb{u}_e}$ and they can be extracted from the derivative $\pd{\mb{E}_e}{\mb{u}_e}$. Now, using Eq.~\ref{eq:straindefinition} and Eq.~\ref{eq:deformationgradient}, we have\footnote{For clarity, the superscript $e$ is left out}
\begin{equation}\label{eq:append_2}
\mb{E} = \frac{1}{2}\left(\nabla_0\mb{u} + \trr{(\nabla_0\mb{u})} + \nabla_0\mb{u}\trr{(\nabla_0\mb{u})}\right),
\end{equation}
In view of FE setting, the displacement vector $\mb{u}$ of an element in terms of its nodal displacements $u_I^A$ and bi-linear shape functions $N_A$ can be written as\footnote{Sum ranges over the number of nodes}: 
\begin{equation}\label{eq:append_3}
\mb{u} = \sum_{A}N_A(\bm{\zeta})u_I^A = N_A(\bm{\zeta})u_I^A.
\end{equation}
Now, Eq.~\eqref{eq:append_2} yields using Eq.~\eqref{eq:append_3} as
\begin{equation}
E_{IJ} = \frac{1}{2}\left(\pd{N_A}{X_J}u_I^A + \pd{N_A}{X_I}u_J^A + \pd{N_A}{X_I}\pd{N_B}{X_J}u_K^A u_K^B\right).
\end{equation}
One finds derivative of $E_{IJ}$ with respect to $u_I^A$ as
\begin{equation}
\pd{E_{IJ}}{u_K^A} = \frac{1}{2}\pd{N_A}{X_J}\left(2\delta_{IK} + 2\pd{N_B}{X_I}u_K^B\right),
\end{equation}
and hence, $\pd{\epsilon_\ms{xx}^e}{\mb{u}_e}$, $\pd{\epsilon_\ms{yy}^e}{\mb{u}_e}$ and $\pd{\epsilon_\ms{xy}^e}{\mb{u}_e}$.

	\end{appendices}
	\bibliography{myreference}
	\bibliographystyle{spbasic} 
\end{document}